\documentclass[10pt,letterpaper]{article}
\usepackage[top=0.85in,left=2.75in,footskip=0.75in]{geometry}

\usepackage{changepage}

\usepackage[utf8]{inputenc}

\usepackage{textcomp,marvosym}

\usepackage{fixltx2e}

\usepackage{amsmath,amssymb}

\usepackage{cite}

\usepackage{nameref,hyperref}

\usepackage[right]{lineno}

\usepackage{microtype}
\DisableLigatures[f]{encoding = *, family = * }

\usepackage{rotating}

\usepackage{float}

\pdfoutput=1 

\raggedright
\usepackage[aboveskip=1pt,labelfont=bf,labelsep=period,justification=raggedright,singlelinecheck=off]{caption}

\usepackage[table]{xcolor}
\usepackage{placeins}

\makeatletter
\renewcommand{\@biblabel}[1]{\quad#1.}
\makeatother

\date{}

\usepackage{lastpage,fancyhdr,graphicx}
\usepackage{epstopdf}
\fancyheadoffset[L]{2.25in}
\fancyfootoffset[L]{2.25in}

\begin{document}
\vspace*{0.35in}

\begin{flushleft}
{\Large
\textbf\newline{Shaping neural circuits by high order synaptic interactions}
}
\newline
\\
Neta Ravid Tannenbaum\textsuperscript{1},
Yoram Burak\textsuperscript{1,2,*},
\\
\bigskip
\bf{1} Edmond and Lily Safra Center for Brain Sciences, Hebrew University, Jerusalem, Israel
\\
\bf{2} Racah Institute of Physics, Hebrew University, Jerusalem, Israel
\\
\bigskip

* yoram.burak@elsc.huji.ac.il

\end{flushleft}
\section*{Abstract}
\noindent Spike timing dependent plasticity (STDP) is believed to play an important
role in shaping the structure of neural circuits. Here we show that STDP generates effective interactions between synapses of different neurons, 
which were neglected in previous theoretical treatments, and
can be described as a sum over contributions from structural motifs. 
These
interactions can have a pivotal influence on the connectivity patterns 
that emerge under the influence of STDP. In particular, we consider 
two highly ordered forms of structure: wide synfire chains, in which groups of neurons project to each other sequentially, 
and self connected assemblies. We show that high order synaptic
interactions can enable the formation of both  
structures, depending on the form of the STDP function and the time course of
synaptic currents. Furthermore, within a certain regime of biophysical parameters,  
emergence of the ordered connectivity occurs robustly and autonomously in a stochastic network of spiking neurons, 
without a need to expose the neural network to structured inputs during learning.  
%

\section*{Author Summary}

Plasticity between neural connections plays a key role in our ability to process and store information. One of the fundamental questions on plasticity, is the extent to which local processes, affecting individual synapses, are responsible for large scale structures of neural connectivity. Here we focus on two types of structures: synfire chains and self connected assemblies. These structures are often proposed as forms of neural connectivity that can support brain functions such as memory
and generation of motor activity. We show that an important plasticity mechanism, spike timing dependent plasticity, can lead to autonomous emergence of these large scale structures in the brain: in contrast to previous theoretical proposals, we show that the emergence can occur autonomously even if instructive signals are not fed into the neural network while its form is shaped by synaptic plasticity.


\section*{Introduction}
One of the striking features of local organisation in the brain is its diversity across brain regions and cell populations \cite{Song2005,Yoshimura2005, Kampa2006}.
Therefore, it is of significant importance to understand the factors that contribute to the organization of local neural circuitry.
Among such factors,
synaptic plasticity involves mechanisms that rely on neural activity 
\cite{Morrison2008,Holtmaat2009}, as first proposed by Hebb \cite{Hebb1949}. 

Spike timing dependent plasticity (STDP), observed in many brain areas, depends on the relative timing of pre and post synaptic spikes
\cite{Markram1997,Bi1998,Froemke2002,Woodin2003}.
This form of plasticity introduces a coupling between the ongoing activity in a neural network and its architecture, because the change in synaptic efficacy is driven by the timing of neural spikes, while the statistics characterizing neural activity are strongly dependent on the connectivity. This link raises interesting questions from a functional and computational perspective.

Considerable theoretical effort was devoted to elucidate the coupling between neural activity and plasticity, when considering modifiable synapses that converge into one post synaptic neuron  \cite{Kempter1999,Gutig2003,Babadi2010,Vogels2011,Luz2012}. When considering STDP in a network of recurrently connected neurons, the coupling between activity and plasticity becomes even more elaborate, since the timing of spikes in each neuron is potentially influenced by the activity of all other neurons in the network. Consequently, the change in one synapse can depend on the full connectivity structure in a complicated manner.

Most 
theoretical works on STDP in recurrent neural networks approximated the dynamics of each synapse as depending on local
quantities: the firing rates of the pre and post synaptic neurons
\cite{Burkitt2007}, and the strengths of the synapses between them \cite{Babadi2013}. 
These local approximations 
provide important insights on synaptic dynamics driven 
by STDP, such as 
the competition that arises between reciprocal 
synapses under an asymmetric STDP function \cite{Song2001,Babadi2013}.
However, 
local approximations do not fully address how plasticity in a given synapse is influenced by other synapses in the network.

Only very recently consequences of STDP in recurrent neural networks
were studied analytically, without resorting to 
the local approximations discussed above. 
In Ref. \cite{Ocker2015} an expression was derived 
for the dynamics of synaptic efficacies in recurrent neural 
networks, based
on an approximation for spike correlations in networks
of leaky integrate and fire neurons. This approach has led to the understanding that STDP
induces effective interactions between pairs of adjacent synapses. These interactions, in turn, 
influence the connectivity, and specifically the 
distribution of local motifs which include pairs of synapses. 
These results highlight the elaborate role played by STDP in shaping the structure of recurrent
neural circuits. 

Many questions remain open, or have received so far only partial answers:
Is it possible to describe in a systematic manner interactions 
between synapses of all orders, beyond 
the pairwise interactions analyzed in previous works?
Can STDP induce global ordered structures in the connectivity, beyond its influence on the
statistical distribution of local motifs within the network? 
How do biophysical properties, such as the structure of the STDP function and the time course of synaptic current, 
influence the emergent structures? 

The first goal of this work is to provide a full description of non-local interactions
between different synapses in recurrent neural networks. 
We develop an expression for the synaptic dynamics in recurrent networks of Poisson spiking neurons, 
which precisely describes how STDP in each synapse depends on the full network connectivity. 
The main advantage of using this neural model is that all the expressions we obtain are 
precise and fully self-consistent, without resorting to any approximations other than those embodied 
in our model for the intrinsic dynamics of individual neurons. 
Furthermore, this approach leads to a systematic 
description of non-local synaptic interactions, expressed as a sum over contributions from structural 
motifs of varying orders.
Using this formalism, we demonstrate that non-local interactions between different synapses
can profoundly influence the synaptic connectivity that emerges under STDP. 
 
\begin{figure}
\begin{centerline}
{\includegraphics[scale=1.0]{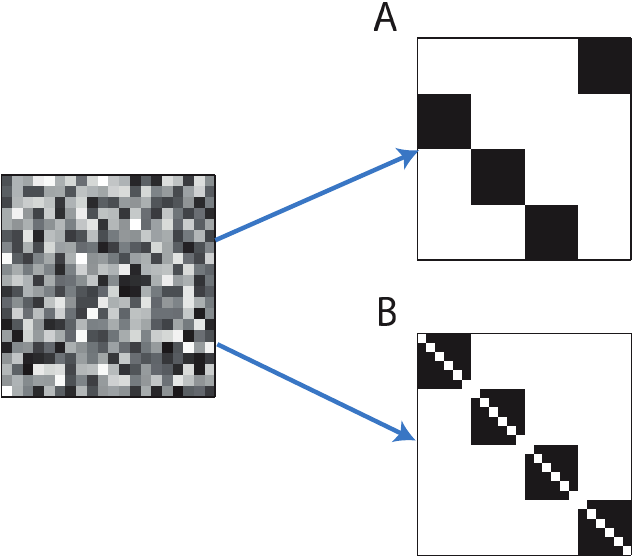}}
\end{centerline}
\protect\caption{\small \textbf{\label{fig:illustration}Spontaneous emergence of global structures from initially random network connectivity. } Illustration of connectivity matrices: Left - random connectivity, \textbf{A}. Synfire chain connectivity, \textbf{B}. Self connected assemblies. In all panels  horizontal (vertical) axes represent the pre (post) synaptic neuron. 
}
\end{figure}

 A second goal of this work is to show that high order, non local interactions between different synapses can promote
 the spontaneous formation of global organization in the connectivity. 
 This result is significant on its own merit (see below), 
 but it also serves as a concrete example for the possible role of high-order synaptic interactions in shaping the structure of neural circuits. 
 We focus on two specific
 types of structures: wide synfire chains, and assemblies of self connected neurons (Fig~\ref{fig:illustration}). 
 The formalism developed here allows us also to predict how biophysical parameters, such as the specific STDP function and the temporal trace of post synaptic currents, impact on the structures that emerge.  
 
\paragraph{Autonomous emergence of global structure.}
Synfire chains, consisting of distinct groups of neurons that project 
to each other in a sequential order (Fig \ref{fig:illustration}A), were originally proposed as a model for sequence generation in the mammalian
cortex \cite{Abeles1991}. Subsequently, compelling evidence 
has pointed to the possibility that 
 this architecture 
 underlies the synchronous neural activity observed in the songbird premotor nucleus 
 HVC \cite{Hahnloser2002,Long2010}, with $\sim$100 excitatory neurons in each
layer. Theoretical reasoning indicates that this architecture can indeed
produce stable propagation of synchronous spiking activity if the groups are sufficiently large  \cite{Diesmann1999,Reyes2003}. 

It was shown theoretically that STDP, combined with heterosynaptic competition, can lead spontaneously to the formation of thin chains \cite{Fiete2010}, in which single neurons 
project to each other sequentially (see also \cite{Buonomano2005}). Formation of wide synfire chains, in which many neurons participate in each group along the sequence, proved to be more
difficult, and was successfully demonstrated  when  structured inputs were fed into the network \cite{Fiete2010}. Thus it remains unclear whether 
structured inputs are required for the 
 formation
of wide chains in networks of spiking neurons.
Here we demonstrate that with appropriate choice of the biophysical parameters, STDP dynamics can lead to robust formation of wide synfire chains, without the need to provide any structured inputs to the network. The grouping of neurons occurs spontaneously, assisted by high order synaptic interactions that arise from the STDP dynamics.

As a second example for the role of high order synaptic interactions in 
plasticity, we consider whether STDP can promote the formation of distinct, self connected groups of cells
(Fig \ref{fig:illustration}B). 
Theoretical works demonstrated that such structures can lead to multiple persistent states
in the neural dynamics \cite{Amit1997,Wang2002,Renart2007,Litwin-Kumar2012,Stern2014}. Another motivation for consideration 
of these structures 
arises from anatomical 
 studies of local connectivity: for example, connections among excitatory neurons
  in the rat visual and somatosensory cortices 
 tend to be clustered \cite{Song2005,berger2011}. 
In the context of learning, it has been demonstrated theoretically that STDP, combined with additional plasticity mechanisms and structured inputs, can lead to formation of self connected assemblies \cite{Amit1997,Clopath2010,Litwin-Kumar2014}. However, in similarity to the formation of synfire chains by STDP, it remains unclear whether self connected assemblies can emerge 
in an initially unstructured network
without the inclusion of correlated inputs that are fed into the network during learning. Here we show that high order synaptic interactions enable the spontaneous formation of self connected assemblies, without the inclusion of such inputs.


\section*{Results}

\subsection*{Synaptic drift in recurrent networks}

\begin{figure}
\begin{centerline}
{\includegraphics[scale=1.0]{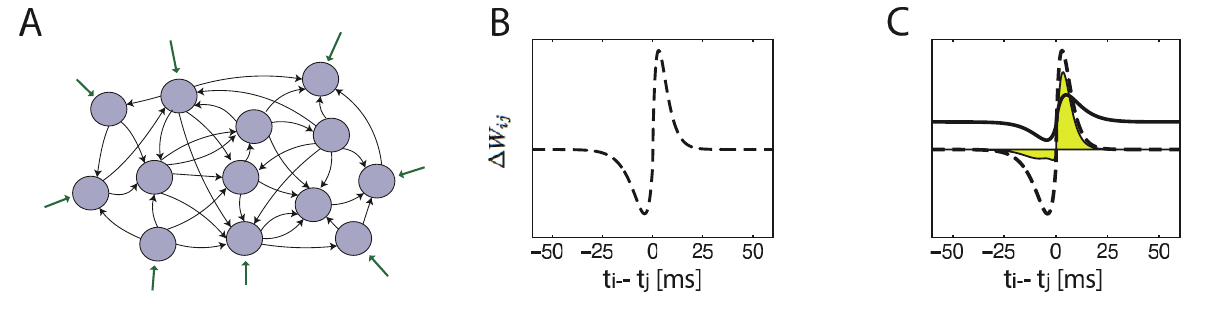}}
\end{centerline}
\vspace*{0.5cm}
\protect\caption{\small \textbf{\label{schematic representation of network}A.}{
Illustration of the network architecture. The network contains recurrently
connected neurons that receive a constant external input. }\textbf{B.
}{Example of an STDP function. Horizontal axis:
time interval between action potentials in the pre and post synaptic neurons.
Vertical axis: change in synaptic efficacy.
}\textbf{C.
}{The drift of the STDP dynamics is an integral over the product of the correlation and STDP functions. Black line: time dependent correlation. Dashed line: STDP function. 
The area under the product of these curves (yellow) determines the synaptic drift.
}}
\end{figure}

In studying the effects of STDP, it is necessary to consider  models of neural activity that explicitly involve the timing of action potentials (as opposed to simpler rate models).
Due to the difficulty in evaluating spike correlation functions in most models of spiking neurons, analytical treatments of STDP have made certain approximations for the spiking statistics: typically, pre and post synaptic spike trains were treated as if they follow inhomogeneous Poisson statistics \cite{Gutig2003,Burkitt2007,Babadi2010,Vogels2011}. 
Therefore, we explicitly consider a recurrent network of neurons which 
follow linear Poisson (LP) dynamics (\textit{Methods}). The activity in the network is stochastic, and the probability of each neuron to emit an action potential is proportional to a weighted sum of the previous activity in the network and a constant external input 
(see Fig~\ref{schematic representation of network}A and  \textit{Methods}).

Networks of LP neurons have been shown to approximate well the correlation in spike timing of neurons with more elaborate leaky integrate-and-fire dynamics, operating in an asynchronous regime \cite{Pernice2012}. The availability of an exact expression for spike correlations in such networks (see below) allows us to develop a precise theory for the weight dynamics driven by STDP.

We consider synaptic efficacies in a recurrent neural network with  an arbitrary structure (Fig~\ref{schematic representation of network}A). The efficacies undergo long term
potentiation or depression in response to each pair of spikes, depending on the time interval between
the firing of the pre and post synaptic neurons (Fig~\ref{schematic representation of network}B). 
In the case of slow learning rate, the rate of change in the synaptic efficacies can be expressed in terms of
the product between the time dependent pair correlation and the STDP function (see \textit{Methods}): 
\begin{equation}
\Delta_{ij}^{{\rm STDP}}\equiv 
\int d\tau \,F\left(\tau\right)C_{ij}\left(\tau\right).
\label{eq: dirft definition}
\end{equation}
Here, $\Delta_{ij}^{{\rm STDP}}$ is the drift in the synaptic efficacy from neuron $j$ to neuron $i$, defined as the average change in the synaptic efficacy per unit time. 
The correlation function $C_{ij}\left(\tau\right)$
represents the probability density that neurons $j$ and $i$ emit a pair of spikes temporally separated by a duration $\tau$ (Eq.~\ref{eq:correlations definition}),
and $F\left(\tau\right)$ is the STDP function that describes how potentiation (or depression) depends on the time interval $\tau$ (Fig~\ref{schematic representation of network}C).

Using an analytic expression for the spike correlation function \cite{The1971}, we obtain an exact expression for STDP in recurrent networks with arbitrary connectivity (\textit{Methods}):
\begin{equation}
\Delta^{\rm {STDP}}
 = f_0 rr^{T}
 + \frac{1}{2\pi}\int_{-\infty}^{\infty}d\omega \tilde{F}\left(-\omega\right)
\left[I-\tilde{a}\left(\omega\right)W\right]^{-1}D
\left[I-\tilde{a}\left(-\omega\right)W^{T}\right]^{-1}\,.
\label{eq: Presize STDP}
\end{equation}
Here, $W$ is the connectivity matrix,  $a(t)$ is the time course of synaptic currents, $\tilde{a}(\omega)$ is its Fourier transform, $r_{i}$ is the average firing rate of neuron $i$ (Eq.~\ref{eq:average firing rate calculation}), $D_{ij}=\delta_{ij}r_{j}$, and $f_0$ is the area under the STDP function (Eq.~\ref{eq:f_0_coefficient}). 
The derivation of Eq.~\ref{eq: Presize STDP} does not involve any assumptions on the specific form of the synaptic currents, STDP function, or the network architecture. 
S1~Fig demonstrates that this expression provides an accurate description of the average learning dynamics in networks of LP neurons, 
over time scales which are relevant to plasticity in the brain.

\subsection*{The synaptic drift can be interpreted as a sum over contributions from structural motifs}

Equation \ref{eq: Presize STDP} expresses how plasticity in one synapse depends on the   connectivity of the full network. Additional insight on this expression, which may seem 
elaborate, is obtained by noticing that the spike correlation functions in the network can be written as a power series, obtained as an
expansion in the strength of the synaptic efficacies 
\cite{Pernice2011}.
This allows us to reformulate Eq.~\ref{eq: Presize STDP} as follows (\textit{Methods}):
\begin{equation}
\Delta_{ij}^{{\rm STDP}}  =f_{0}r_{i}r_{j}+  \sum_{\alpha\beta}f_{\alpha,\beta}\cdot\sum_{k}r_{k}\left(W^{\alpha}\right)_{ik}\left(W^{\beta}\right)_{jk},
\label{eq:power decomposition}
\end{equation}
where the coefficients $f_{\alpha, \beta}$ are defined below. Each one of the terms in Eq.~\ref{eq:power decomposition} has a relatively simple, intuitive interpretation that we discuss next.

\begin{figure}
	\begin{centerline}
	{\includegraphics[scale=0.65,trim=3.5cm 0cm 0cm 0cm, clip=true]{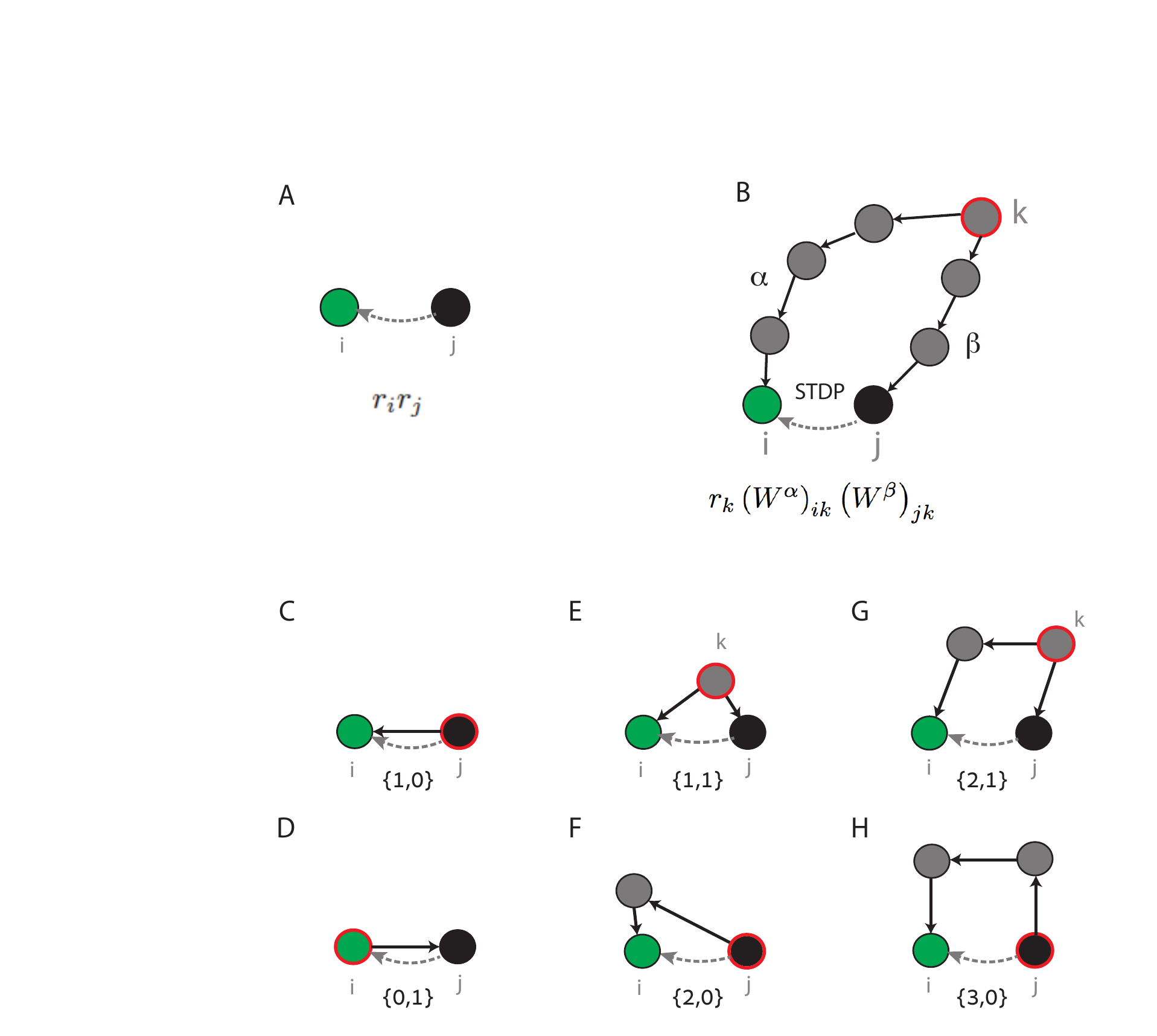}}
	\end{centerline}
	\vspace*{0.5cm}
	\protect\caption{\small
		\textbf{\label{fig:Network-motifs}Network motifs that affect the STDP dynamics.}
		\textbf{A. } 
		The first term in Eq.~\ref{eq:power decomposition} is proportional to the average firing 
		rates of the pre and post synaptic neurons.  
		\textbf{B. }General form of a structural motif, 
		contributing to the STDP dynamics in the synapse $W_{ij}$ (dashed line).  
		A source neuron $k$ (red line) projects to the post synaptic neuron $i$ (green) via $\alpha$ synapses, and to the pre synaptic neuron $j$ (black) via $\beta$ synapses. The contribution is proportional to the firing rate of the source neuron $r_{k}$ and to the product of all synaptic efficacies along the two paths. The expression $\left(W^{\alpha}\right)_{ik}r_{k}\left(W^{\beta}\right)_{jk}$ is a sum over all paths that start from neuron $k$ and reach the post synaptic neuron $i$ via $\alpha$ synapses and the pre synaptic neuron $j$ via $\beta$ synapses. 
		\textbf{C-H:}
		Examples of six motifs that affect STDP in the synapse $W_{ij}$
		(dashed line), by inducing time dependent coupling between neurons $j$ 
		and $i$. 
		Black arrows indicate
		synapses that participate in the motifs. \textbf{C.}
		One of the first order motifs contains the direct synapse from $j$ to $i$ ($\alpha=1$, $\beta=0$). The source neuron is the pre synaptic neuron $j$. \textbf{D. } The
		other first order motif contains the opposite synapse from $i$ to $j$ ($\alpha=0$, $\beta=1$). Here the source neuron is the post synaptic neuron $i$. \textbf{E.} 
		One of the second order motifs ($\alpha=1$, $\beta=1$). Here
		the source neuron is some other neuron in the network that projects
		directly both to $i$ and $j$. \textbf{F. }
		Additional example for a second order motif ($\alpha=2$, $\beta=0$). The source neuron is the
		pre synaptic neuron $j$, that projects to the post synaptic neuron
		$i$ through one intermediate neuron. \textbf{G.}
		Example of a third order  motif ($\alpha=2$, $\beta=1$). 
		The source neuron projects directly to the pre synaptic neuron $j$ 
		and via one intermediate neuron to the post synaptic neuron $i$. 
		\textbf{H.  }Additional example of a third order motif ($\alpha=3$, $\beta=0$). The pre synaptic neuron $j$ is the source neuron, that projects to the post synaptic neuron $i$
		via two intermediate neurons.  
	}
\end{figure}

The first term in Eq.~\ref{eq:power decomposition}, $f_{0}r_{i}r_{j}$, represents the contribution to STDP arising from the mean firing rates of the pre and post synaptic neurons, while ignoring any correlations in the timing of their spikes. 
Accordingly, this term is simply proportional to the firing rates $r_i$ and $r_j$ (Fig~\ref{fig:Network-motifs}A). 
Such a term is often postulated in phenomenological models of synaptic plasticity \cite{Linsker1986}, and its emergence from STDP dynamics has been described, e.g., in \cite{Kempter1999,Gutig2003,Burkitt2007,Babadi2013}. 

The probability of neurons $i$ and $j$ to emit a spike is transiently modulated whenever a spike is emitted anywhere within the network.
In the sum on the right hand side of 
Eq.~\ref{eq:power decomposition},
each term quantifies how a spike in a neuron $k$ modulates
the probability of neurons $i$ and $j$ to emit pairs of spikes at various latencies -- and through this modulation, how the
spikes of neuron $k$ influence the drift in the synaptic efficacy $W_{ij}$. This contribution to the drift is written as a sum over structural motifs, which share a common organization shown schematically in Fig~\ref{fig:Network-motifs}B.

In each structural motif, the source neuron $k$ projects to the post synaptic neuron $i$ via a path of $\alpha$ synapses, and to the pre synaptic neuron $j$ via a path of $\beta$ synapses. The synaptic drift driven by the motif
is proportional to all the synaptic weights along the two paths, and to the firing rate of the source neuron. In addition, the drift is proportional to a \textit{motif  coefficient} 
$f_{\alpha, \beta}$. This coefficient depends on the number of synapses in the two paths, 
the time course of the synaptic currents, and 
the detailed form of the STDP learning function. We discuss this dependence in detail later (see also \textit{Methods}). 

The first order contributions in the above sum are those in which $\{\alpha,\beta\} = \{1,0\}$ (Fig~\ref{fig:Network-motifs}C), or $\{\alpha,\beta\} = \{0,1\}$ (Fig~\ref{fig:Network-motifs}D): 
\begin{equation}
\Delta_{ij}^{{\rm STDP}}=f_{0} r_{i}r_{j}+f_{1,0} r_{j}W_{ij}+f_{0,1} r_{i}W_{ji}+\ldots\,.
\end{equation}
These terms are local: they depend only on the direct synapses that link neurons $i$ and $j$, and on the firing rate of these two neurons (Fig~\ref{fig:Network-motifs}C,D). 
Previous works \cite{Song2001,Babadi2013} derived
these contributions to STDP using heuristic arguments that focused on the pre and post synaptic neurons, and studied their consequences when embedded in a recurrent network.
Under an asymmetric STDP function, the first order terms induce a competition between a synapse $W_{ij}$ and the opposite synapse $W_{ji}$ \cite{Song2001,Babadi2013}, whereas a symmetric STDP function tends to promote the development of a symmetric weight matrix. Here, these local plasticity rules are obtained as the first order terms in a systematic expansion, which includes also higher order terms. 

\paragraph{Synapses of different neurons affect each other through higher order terms}
A central consequence of Eq.~\ref{eq:power decomposition}
is that the drift in one synapse can be affected by other synapses in the network through the contribution of high order motifs. 
An illustration of how high order motifs induce interactions between different synapses is seen in two examples of second order motifs, shown in Fig~\ref{fig:Network-motifs}E,F. 
In Fig~\ref{fig:Network-motifs}E, the source neuron $k$ is an arbitrary neuron in the network, that projects directly to neurons $j$ and $i$. STDP in $W_{ij}$ is influenced, through this motif, by the synaptic efficacies $W_{jk}$ and $W_{ik}$. 

In the motif shown in Fig~\ref{fig:Network-motifs}F, the source neuron is the neuron $j$ itself, and the motif includes a path leading from neuron $j$ to neuron $i$.
Here the effect of a spike in neuron $j$ on neuron $i$ is mediated through an intermediate neuron. Thus, the synapse $W_{ij}$ is influenced, via this motif, by synapses along indirect paths connecting neuron $j$ and $i$ through a single intermediate neuron.
Additional examples, of third order motifs, are shown in Fig~\ref{fig:Network-motifs}G,H. 

\subsection*{High order motifs can promote self organization into global structures}

\begin{figure}
\begin{centerline}
{\includegraphics[scale=0.7]{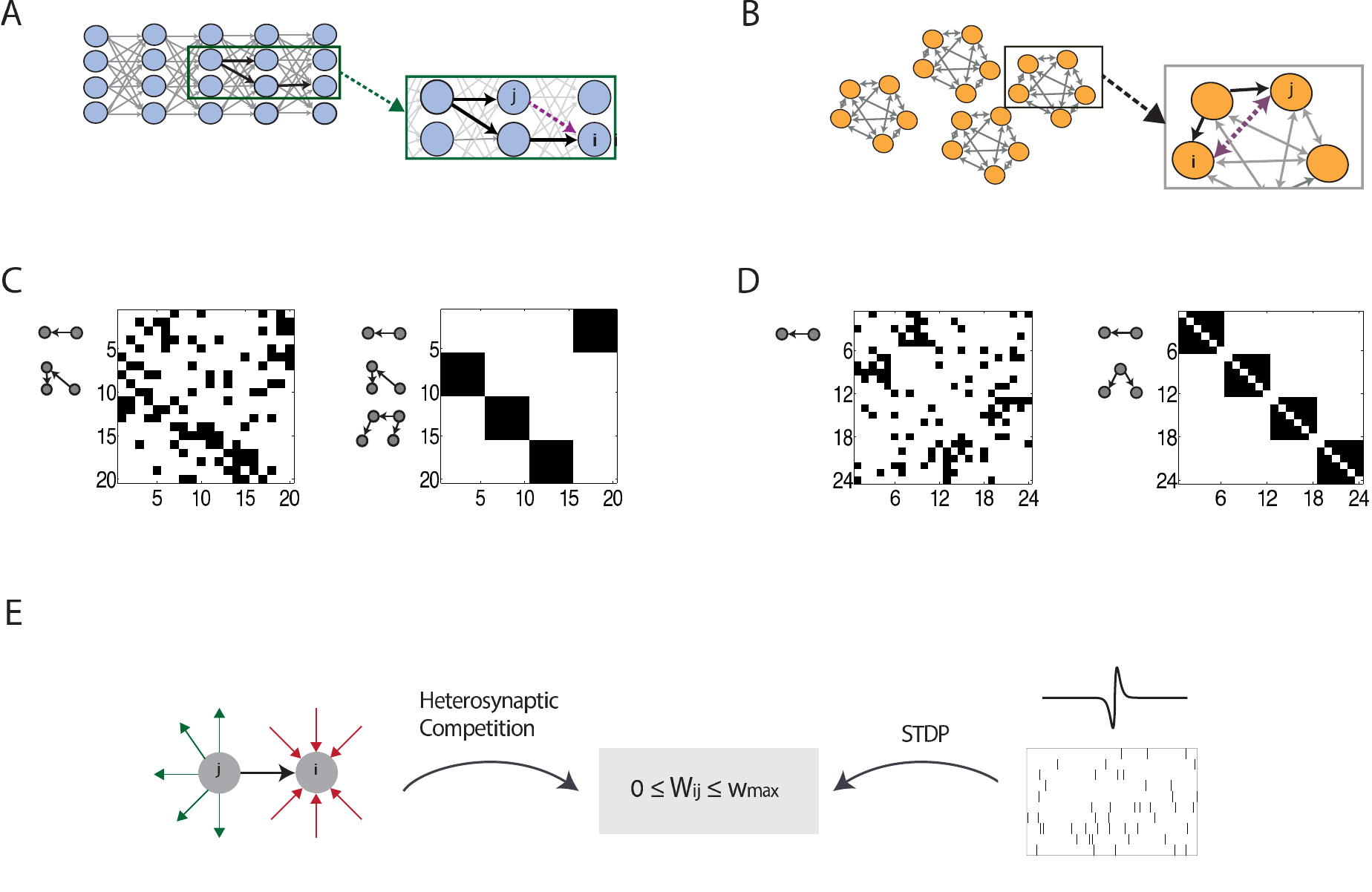}}
\end{centerline}
\vspace*{0.5cm}
\protect\caption{\small
\textbf{\small \label{fig:complex structures}Self organization into global structures}.
 \textbf{A. }{ The third order motif $\{2,1\}$ promotes formation of wide synfire chain connectivity, 
 by encouraging neurons that receive input from the 
same pre synaptic neuron to project to the same post synaptic neuron. 
}\textbf{B.}{ 
The second order motif $\{1,1\}$, combined with a symmetric STDP function, promotes formation of self connected assemblies by
encouraging two neurons that receive inputs from a common source to increase their synaptic coupling reciprocally. 
}\textbf{C-D. }{ Steady state connectivity, obtained from simulations in which the motif coefficients $f_{\alpha,\beta}$ are tuned artificially. STDP is combined with heterosynaptic competition. Neurons are ordered such that the matrix represents the optimal feed forward connectivity (C) or self connected assemblies connectivity (D). 
}\textbf{C. }{ Left - results from dynamics that contain only the motifs $\{1,0\}$, $\{0,1\}$, $\{2,0\}$, $\{0,2\}$. The simulation does not converge into synfire chains connectivity. Right - the simulation contains also the motifs $\{2,1\}$, $\{1,2\}$. The simulation converges into synfire chain connectivity. 
}\textbf{D. }{Left - the dynamics contain only the motifs $\{1,0\}$, $\{0,1\}$. The simulation does not converge to assemblies. Right - the dynamics include also the motif $\{1,1\}$. The connectivity converges into self connected assemblies.
} \textbf{E.  }{The drift in the excitatory connections is driven by three plasticity mechanisms: (i) Spike timing dependent plasticity. (ii) Heterosynaptic competition between all synapses that terminate in the same post synaptic neuron (red arrows), and all synapses that originate from the same pre synaptic neuron (green arrows). (iii)
A hard bound on the synaptic efficacy.}
}
\end{figure}

Next, we demonstrate
that high order motifs can promote the formation of large-scale structures in the synaptic connectivity. We focus on two types of structures: synfire chains (Fig~\ref{fig:complex structures}A), and clusters of self connected assemblies (Fig~\ref{fig:complex structures}B). In both structures, synapses of different neurons are highly correlated. 
The purpose of this section is to illustrate 
by specific examples that high order motifs, beyond the first order, can lead to emergence of these
structural correlations. A more 
systematic study is presented in later sections. 

We consider networks that consist of recurrently connected excitatory neurons, and inhibitory interneurons with fast synapses (see \textit{Methods}). 
The inhibitory input to each neuron depends on the total activity of the excitatory network, and is adjusted such that the inhibitory weights balance the excitatory ones  \cite{Vogels2011,Luz2012}. The main role of inhibition in the network is to
suppress runaway excitation in the neural dynamics.
In addition, all neurons receive constant, identical inputs (\textit{Methods}). 

The plasticity mechanisms acting on the excitatory neurons are summarized in Fig~\ref{fig:complex structures}E.
The excitatory connections are modifiable through STDP and are bounded between zero and a positive bound, denoted by $w_{\rm{max}}$. In addition to STDP, the excitatory synapses undergo heterosynaptic competition that limits the total synaptic input and output of each neuron: 
the sum of the incoming excitatory weights to each neuron, and the sum of outgoing excitatory weights from each neuron are bounded to lie below a positive hard bound, denoted by $W_{\rm max}$. The competition, combined with STDP and with the hard bound on each synaptic weight, can lead to a steady state connectivity pattern in which each neuron receives input from a certain number of pre-synaptic partners, and projects to a certain number of post-synaptic partners \cite{Fiete2010}. These numbers are tuned by the ratio between $W_{\rm max}$ and
$w_{\rm max}$ \cite{Fiete2010}.

To test the influence of individual motifs on the 
STDP dynamics, we perform simulations in which we include only a few of the terms in Eq.~\ref{eq:power decomposition},
starting from initial weights that were drawn independently from a uniform random distribution (\textit{Methods}).
Instead of 
obtaining the exact expressions for the coefficients $f_{\alpha,\beta}$ (Eq.~\ref{eq:motif coefficients}), we 
artificially tune their values and observe
the consequences on the structures that emerge. 

The motif $\{2,1\}$ (Fig~\ref{fig:Network-motifs}G), when acting between excitatory neurons, 
encourages neurons that receive input from the same pre synaptic
neuron to project into the same post synaptic neuron (Fig~\ref{fig:complex structures}A).
Consequently, this motif can induce correlations between the synaptic connections formed by
neurons that belong to the same layer of a synfire chain. 
This is a key feature which differentiates
wide synfire chain structures from other connectivity patterns in which each neuron has a prescribed number 
of presynaptic and postsynaptic partners. 
This observation raises a hypothesis, that the motif 
$\{2,1\}$ can promote formation of wide synfire chains,
by favoring these structures over other connectivity patterns which are compatible with 
the constraints set by the heterosynaptic competition. 

To test this hypothesis, we perform simulations that
include only contributions from the motifs
$\{\alpha, \beta\} =\{1,0\}$, $\{2,0\}$,
$\{2,1\}$, and the opposing terms in which $\alpha$ and $\beta$ are exchanged. Additionally, we set  
$f_{\alpha, \beta} = -f_{\beta, \alpha}$, as expected if the STDP 
function is 
antisymmetric. 

An example is shown in Fig~\ref{fig:complex structures}C. When including only the first and second order motifs, a simulation of the plasticity dynamics leads to a structure in which each row and column contains a small number of active synapses, without any reciprocal connections (left). However, the synaptic weights are not organized in a synfire chain structure. When including the third order motifs $\{2,1\}$ and $\{1,2\}$, the synaptic efficacies self organize into a perfect synfire chain (right), despite the absence of any correlations in the external inputs to the network. 
Results from a wider set of simulations, in which we systematically vary 
the strength of motifs, are presented in the section \textit{Self organization into synfire chains}. 

As a second example, we examine the influence of the motif $\{1,1\}$ on the formation of 
self connected cell assemblies. 
The motif $\{1,1\}$ (Fig~\ref{fig:Network-motifs}E), when acting between excitatory neurons, 
enhances reciprocal connections between neurons that receive common input 
(note that the contribution from this motif vanishes if the STDP function is antisymmetric, but for 
a symmetric STDP function this motif can significantly contribute to the plasticity dynamics). 
This raises the hypothesis 
that formation of self connected assemblies can be promoted by the contribution of 
the motif $\{1,1\}$.

To check this hypothesis, we conduct plasticity simulations that contain only 
contributions from first and second order motifs:
$\{0,1\}$, $\{1,0\}$, and $\{1,1\}$ in 
 Eq.~\ref{eq:power decomposition}. In addition, we impose the relation $f_{1,0}=f_{0,1}$, 
as expected if the STDP function is symmetric (see Eq.~\ref{eq:motif coefficients} in \textit{Methods}).
In the example shown in Fig~\ref{fig:complex structures}D, the first order motifs, together with the synaptic competition, lead to a symmetric connectivity matrix in which the number of active synapses is limited in each row and in each column (left). With the inclusion of the $\{1,1\}$ motif (right), strong correlations emerge between synapses of different neurons, and fully connected cell assemblies emerge. 

\subsection*{Biophysical properties affect the relative contribution of different motifs to STDP}

So far, we illustrated how high order motifs can promote the formation of global structures 
by artificially tuning the contribution of specific motifs to  plasticity (through the coefficients $f_{\alpha,\beta}$). 
In the actual STDP dynamics, the coefficients $f_{\alpha,\beta}$ cannot be controlled independently. Instead, these coefficients are determined by the
temporal structure of the STDP function and the synaptic currents. 

Each motif induces temporal correlations of spikes between the pre and post synaptic neurons with a characteristic
time course. The time course of correlations depends on the structure of the motif (characterized by $\alpha$ and $\beta$), 
and on the time course of the synaptic currents. 
Therefore, the synaptic current, together with the STDP function, affects how strongly each motif
 influences the synaptic drift, as
 quantified by the magnitude of the corresponding motif coefficient $f_{\alpha, \beta}$ (Eq.~\ref{eq:motif coefficients}). 
The influence of the synaptic currents on the motif coefficients is illustrated 
in detail in \textit{Methods}.

\begin{figure}
	\begin{centerline}
	{\includegraphics[scale=0.7]{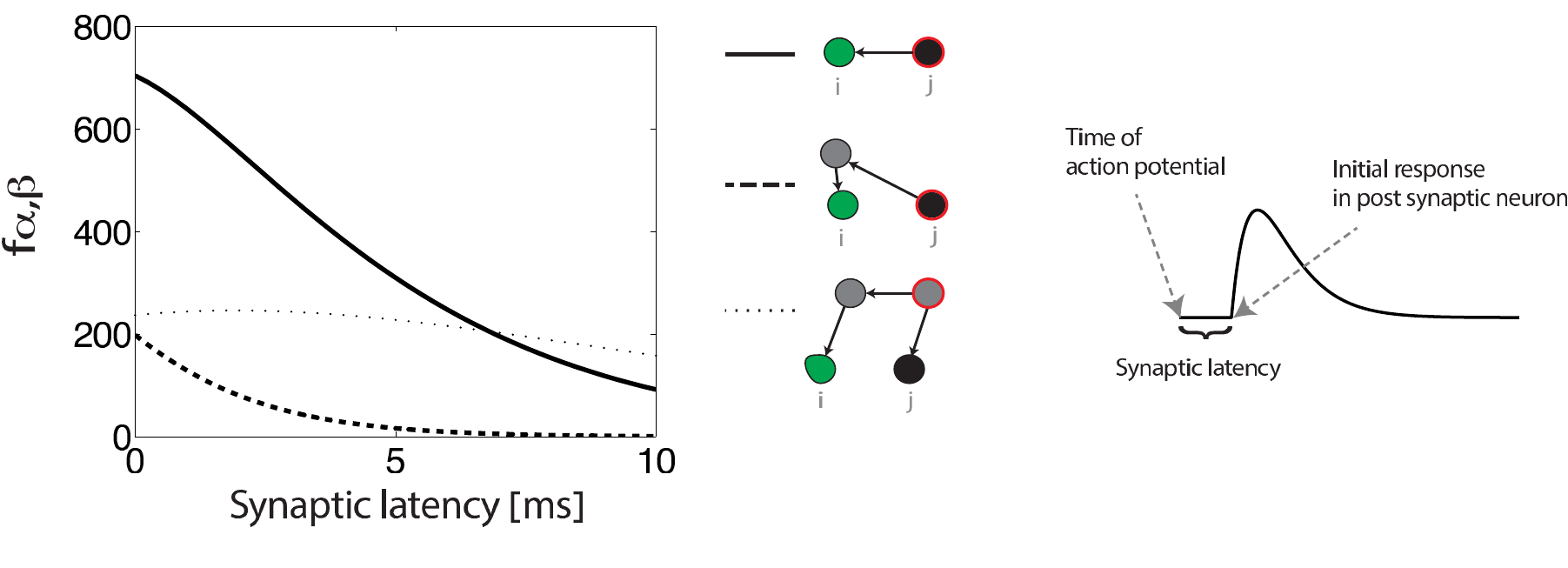}}
	\end{centerline}
	\vspace*{0.5cm}
	\protect\caption{\small \textbf{\label{fig:-Motifs coefficients} 
		}{ Left: Motif  coefficients as a function of
		the synaptic latency. Solid line - $f_{1,0}$ , dashed line - $f_{2,0}$, dotted line - $f_{2,1}$.
		An antisymmetric STDP function is assumed. Right: the time course of the synaptic current. 
		The synaptic latency is modeled as a delay in the onset of the postsynaptic current, 
		relative to the presynaptic spike time. The precise form of the synaptic current and STDP function 
		are specified in \textit{Methods}.}}
\end{figure}

As a specific example for how the time course of synaptic currents can affect the motif coefficients,
we consider in Fig~\ref{fig:-Motifs coefficients} the 
influence of a delay in the onset of the post synaptic current
(abbreviated below as the \textit{synaptic latency}). 
Synaptic latencies
depend on diverse physiological properties, 
such as the length and conductance velocity in the axon \cite{Sabatini1999} and location on the dendrite \cite{Rall}. 
Here we model the synaptic latency as a temporal shift in the synaptic current
(Fig~\ref{fig:-Motifs coefficients}).

In the example shown in Fig~\ref{fig:-Motifs coefficients}
the motif coefficients $f_{1,0}$ and $f_{2,0}$
decrease with an increase of the synaptic latency. The coefficient $f_{2,1}$ is influenced by the synaptic latency as well,
but for synaptic latencies ranging from 0 to 10\,ms this dependence is extremely weak.
Therefore, 
an increase in the synaptic latency reduces the contribution of the motifs $\{1,0\}$ and $\{2,0\}$ to STDP relative to the motif
$\{2,1\}$. The reason for these trends is explained qualitatively in \textit{Methods}.
Higher order motifs exhibit a similar behavior, depending on the difference between $\alpha$ and $\beta$ (S5~Fig).

The possibility to tune the relative contribution of motifs through the interplay of post synaptic currents and the STDP function, suggests that global structures could be spontaneously generated via STDP with appropriate choice of these biophysical parameters. Furthermore, this observation provides a principled way to search for parameters 
that enable emergence of specific structures.

\subsection*{Self organization into synfire chains}

We next focus on the emergence of synfire chains under the influence of STDP
and heterosynaptic competition.
For simplicity, we consider mainly the case where the STDP function is antisymmetric. 

\begin{figure}
	\begin{centerline}
	{\includegraphics[scale=1.0]{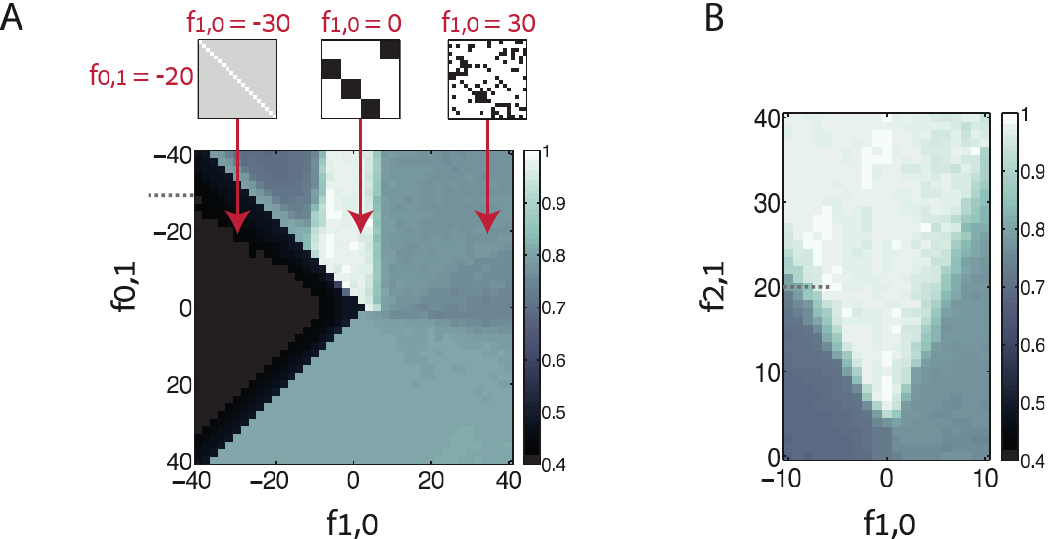}}
	\end{centerline}
	\protect\caption{\small \textbf{\label{fig:Synfire-chains motifs}Chain score of the steady state connectivity, obtained from simulations that include selected motifs. A. }{Density
			plot of the chain score, displayed as a function of the motif coefficients $f_{1,0}$ (horizontal axis) and $f_{0,1}$ (vertical axis). All other motif coefficients are kept fixed. 
			Note that here $f_{2,1} = 20$ (designated by a dotted line in panel B).
			Above: examples of steady state connectivity matrices for $f_{0,1}=-20$, and (from left to right) $f_{1,0}=-30,\rm{ } 0,\rm{ } 30$. 
		}\textbf{B.}
		Chain score as a function of the motif coefficients $f_{1,0}$ (horizontal axis) and $f_{2,1}$ (vertical axis) with all other motif coefficients kept fixed. 
		Here $f_{0,1} = -30$ (designated by a dotted line in panel A).
		Each data point represents an average over ten simulations, each with a different realization of the initial random connectivity. 
		Additional details of the plasticity dynamics are
		specified in \textit{Methods}.
	}
\end{figure}

First, we consider in detail the interplay between the third order motif $\{2,1\}$, which facilitates 
the formation of synfire chains,
and the first order motifs $\{1,0\}$ and $\{0,1\}$, whose contribution to STDP dynamics was the focus of previous theoretical works \cite{Babadi2013,Song2001}. 
Intrinsic plasticity mechanisms other than STDP can act to self-regulate the efficacy of each synapse, 
in similarity to the effect of the
first order motif $\{1,0\}$ (see below). Therefore, it is interesting to consider $f_{1,0}$ and $f_{0,1}$ as separate parameters 
even if the STDP function is antisymmetric.
In Fig~\ref{fig:Synfire-chains motifs} we examine the phase space spanned by
$f_{1,0}$, $f_{0,1}$, and $f_{2,1}$, while assuming that $f_{1,2} = -f_{2,1}$ due to the antisymmetric form of the STDP function. 
To avoid decay of all weights to zero when $f_{1,0}$ is strongly negative, we include in the dynamics also a term that drives
growth of each weight at a fixed rate (see \textit{Methods}, 
Eq.~\ref{eq:Complete learning rule}).
For simplicity,
we first consider a situation in which other motifs do not contribute to the dynamics.

Fig~\ref{fig:Synfire-chains motifs}A shows results from simulations in which 
we varied $f_{0,1}$ and $f_{1,0}$
while fixing $f_{2,1}$.
To quantify whether synfire chains emerge robustly we constructed 
a score that quantifies similarity between the steady state excitatory 
connectivity and a perfect synfire chain structure
(abbreviated below as the \textit{chain score}). 
This chain score ranges from 0 to 1, where 1 corresponds to a perfect match
(\textit{Methods}). 
The figure shows the chain score, averaged over multiple random choices of the initial weights. 
Over a wide range of values of $f_{1,0}$ and $f_{0,1}$ 
precise synfire chain structures are obtained robustly. 
We note two characteristics of this parameter regime: 
first, the coefficient $f_{0,1}$ is negative. Thus, each synapse inhibits its reciprocal synapse.
Second, $f_{1,0}$ lies within a range of values which is fairly insensitive to $f_{0,1}$ when 
$\left|f_{0,1}\right|$ is sufficiently large. 

Similarly, in Fig~\ref{fig:Synfire-chains motifs}B we fix $f_{0,1}$, 
while varying $f_{1,0}$ and $f_{2,1}$.
As expected, synfire chains emerge only when $f_{2,1}$ is sufficiently large. 
Furthermore, with increase of $f_{2,1}$ the range of 
values of $f_{1,0}$ that permits formation of synfire chains becomes wider.
Thus, the high order motif $\{2,1\}$ plays a pivotal role in the spontaneous emergence
of wide synfire chains. 
Results from additional simulations, in which we include additional low order motifs are shown in 
S2~Fig. Under typical conditions relevant to the full plasticity dynamics (discussed below),
the second order 
motifs $\{2,0\}$ and $\{1,1\}$ have a detrimental
effect on synfire chain formation
(a contribution from the motif $\{1,1\}$ may be present if the STDP function is not antisymmetric).
Next, we address the emergence of synfire chains under the full STDP dynamics. 

\paragraph{Synaptic self-depression enables self-organization into synfire chains in the full STDP dynamics}

With typical choices of the STDP function and the synaptic currents,
the contribution of the first order motif $\{1,0\}$ to the dynamics is large relative to
higher order motifs (Fig~\ref{fig:-Motifs coefficients}). 
Consequently,
based on the results shown in Fig~\ref{fig:Synfire-chains motifs} 
we do not expect emergence of synfire chain structures under STDP dynamics alone. 
To enable formation of synfire chains, we include in the 
dynamics a term that describes constant self depression of each excitatory synapse, which acts
alongside STDP (see \textit{Methods}). 
Such a term can arise, 
for example, from a self-regulating process which decreases the size of each dendritic spine in proportion to its volume \cite{Loewenstein2011}. 
Hence, we assume that the self depression term, acting on each synapse
 is proportional to the synaptic efficacy ($-\mu W_{ij}$ in Eq.~\ref{eq:Complete learning rule}).
 This form of 
synaptic self-depression 
selectively suppresses the contribution of the motif $\{1,0\}$, since its contribution
is similar to that of the motif $\{1,0\}$.
An increase in the rate of self-depression corresponds  in 
Fig~\ref{fig:Synfire-chains motifs}A to motion from right to left, in parallel to the horizontal axis. In light of 
Fig~\ref{fig:Synfire-chains motifs} we expect synfire chains to emerge for appropriate rates of the synaptic self-depression.

\begin{figure}
	\begin{centerline}
	{\includegraphics[scale=0.8]{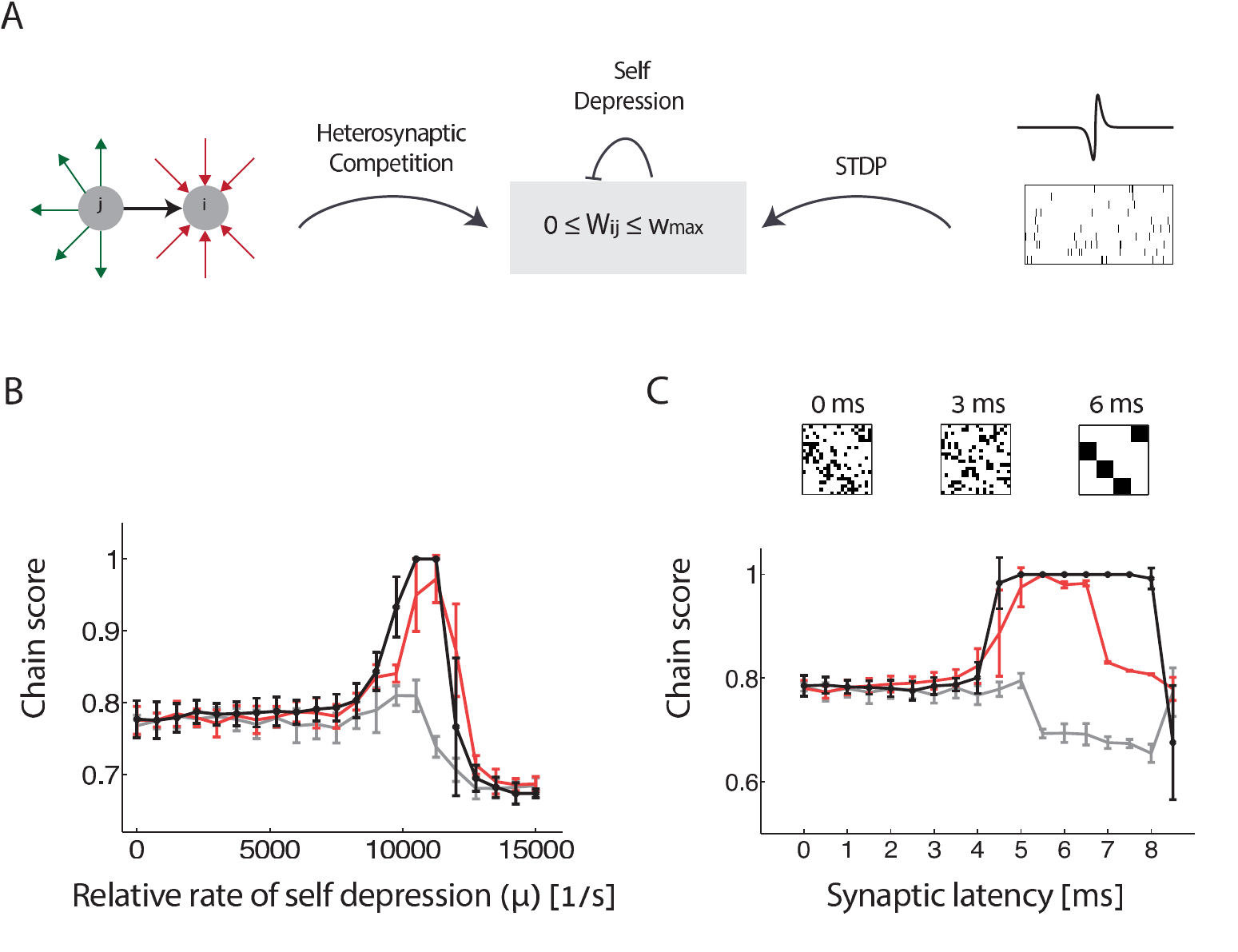}}
	\end{centerline}
	\protect\caption{\small \textbf{\label{fig:Synfire-chains _full_dynamics}Emergence of synfire chains under the full STDP dynamics. A. }
		{Schematic
			representation of the plasticity rules applied to excitatory synapses. In addition to the mechanisms mentioned in 
			Fig~\ref{fig:complex structures}, the synapses undergo self depression that weakens their efficacy in proportion to its magnitude.}
		\textbf{B-C.
		}
		{Chain score of the steady state connectivity, averaged over ten random choices of the initial connectivity. 
			Black traces: complete dynamics (Eq.~\ref{eq: Presize STDP}); red (gray) traces - power series expansion (Eq.~\ref{eq:power decomposition}), truncated to include motifs up to third (second) order. 
			\textbf{B.} Chain score plotted as a function of $\mu$, the relative rate of self-depression (the rate of self depression is given by $\eta \mu$, where $\eta$ is the learning rate, see \textit{Methods}).  
			All other parameters are kept fixed. \textbf{C}. Chain score plotted as a function of the synaptic latency, while keeping all other parameters fixed. 
			Above: three examples of the steady state connectivity obtained in specific simulations, with different values of synaptic latency. Error bars in panels B-C represent the standard deviation of the
			chain score over the ten initial conditions. 
		}}
	\end{figure}

The black trace in Fig~\ref{fig:Synfire-chains _full_dynamics} shows results from simulations of the full STDP dynamics
(Eq.~\ref{eq: Presize STDP}) combined with synaptic self-depression and
heterosynaptic competition (Eq.~\ref{eq:Complete learning rule}). 
All the processes contributing to the synaptic dynamics are summarized schematically in 
Fig~\ref{fig:Synfire-chains _full_dynamics}A. In Fig~\ref{fig:Synfire-chains _full_dynamics}B
chain scores of the steady state connectivity, averaged over multiple simulations with random initial synaptic weights, 
are shown as a function of $\mu$, the rate of self depression.
All other parameters in the simulation are kept fixed. As expected, synfire chains emerge robustly
when $\mu$ lies within an appropriate range. 

The red trace in Fig~\ref{fig:Synfire-chains _full_dynamics}B shows similar results 
from simulations in which the expansion of the STDP dynamics 
(Eq.~\ref{eq:power decomposition}) was truncated to include only contributions of motifs up to third order. 
Comparison with results of the full STDP dynamics (solid line) indicates 
that the contributions of motifs up to third order are
sufficient to predict quite well the conditions in which
synfire chains emerge. In contrast, synfire chain structures do not emerge when the expansion in Eq.~\ref{eq:power decomposition} is truncated
to include only motifs up to second order (Fig~\ref{fig:Synfire-chains _full_dynamics}B, gray trace). 

\paragraph{Synaptic latency can facilitate self organization into synfire chains}
Figure~\ref{fig:-Motifs coefficients} demonstrates that a synaptic latency of several
ms
reduces the motif coefficients $f_{1,0}$ and $f_{2,0}$, 
while having only little influence on the third order motif coefficient 
$f_{2,1}$. Therefore, we consider how a synaptic latency influences the
emergence of synfire chain structures.

Fig~\ref{fig:Synfire-chains _full_dynamics}C shows that in the full STDP dynamics,
perfect synfire chain structures emerge robustly within a certain range of synaptic latencies 
(black trace).
Several examples of steady state structures, 
obtained with different synaptic latencies are shown at the top of panel C.
Qualitatively, varying the synaptic latency has a similar effect as that of self-depression, 
since both mechanisms decrease the contribution of the motif $\{1,0\}$, with little or no 
effect on the contribution of the third order motif $\{2,1\}$.
However, the influence of the synaptic latency is more elaborate
than that of self-depression, since the latter mechanism 
tunes only the contribution of the motif $\{1,0\}$, whereas the synaptic latency influences
(in general) all the motif
coefficients. Thus, 
varying the synaptic latency traces a curve within 
the phase space of motif coefficients in which 
most motif coefficients vary.
The red trace in Fig~\ref{fig:Synfire-chains _full_dynamics}C demonstrates 
that even in this more elaborate situation, it is sufficient 
to include contributions of motifs
up to third order in order to qualitatively predict the influence of the synaptic latency in 
the full STDP dynamics.

In S3~Fig we consider 
the influence of the strength of synaptic weights. An increase in the synaptic weights
amplifies the relative contribution of high-order motifs (Eq.~\ref{eq:power decomposition}). 
As expected, the outcome is a
widening of the permissive range for synfire formation.
Finally, in S4~Fig we demonstrate that synfire chain structures can emerge robustly also for an STDP function 
which is not precisely
antisymmetric. Thus, the $\{1,1\}$ motif contributes to the dynamics, and so do other
high order motifs with $\alpha=\beta$. We assumed that the area under the STDP function is positive:
in this case it is possible to set $\gamma=0$ (Eq.~\ref{eq:Complete learning rule}), because
the zeroth order term of the dynamics (Eq.~\ref{eq:power decomposition}) is 
sufficient to drive growth of all the synapses.

\paragraph{Robustness to noise}
The average synaptic drift as expressed by Eq.~\ref{eq: Presize STDP}
is sufficient to describe the plasticity dynamics when the learning rate is small and noise in the 
STDP dynamics, arising from random fluctuations in the number of pre and post spike pairs is averaged out. 
Thus, the analysis in the limit of slow learning does not guarantee that the network will exhibit similar plasticity dynamics in
a more realistic scenario, where
learning occurs over a biologically relevant time scale. 

A rough estimate for the dependence of noise on the time scale of averaging
can be obtained by comparing the 
prediction of the deterministic theory (Eq.~\ref{eq: Presize STDP}) 
with the synaptic efficacy generated in a stochastic spiking simulation. Such results are shown in S1~Fig, 
where the network parameters were chosen to roughly match those used in Fig~\ref{fig:Synfire-chains _full_dynamics}. 
The synaptic drift  
is correctly predicted by Eq.~\ref{eq: Presize STDP} 
and, as expected, the  
scatter relative to the predicted drift decreases with increase of the duration of averaging $T$. 
When averaged over
a fairly short time scale of $T = 6\,$min, the stochastic drift is strongly correlated with the prediction
of the deterministic theory, but the scatter relative to the mean is fairly large
(S1~Fig\,A). When the duration of averaging is 200 times longer ($T = 20$\,hours) the scatter is much smaller, and the
agreement with Eq.~\ref{eq: Presize STDP} is excellent (S1~Fig\,C). 

\begin{figure}
	\begin{centerline}
{\includegraphics[scale=1.0]{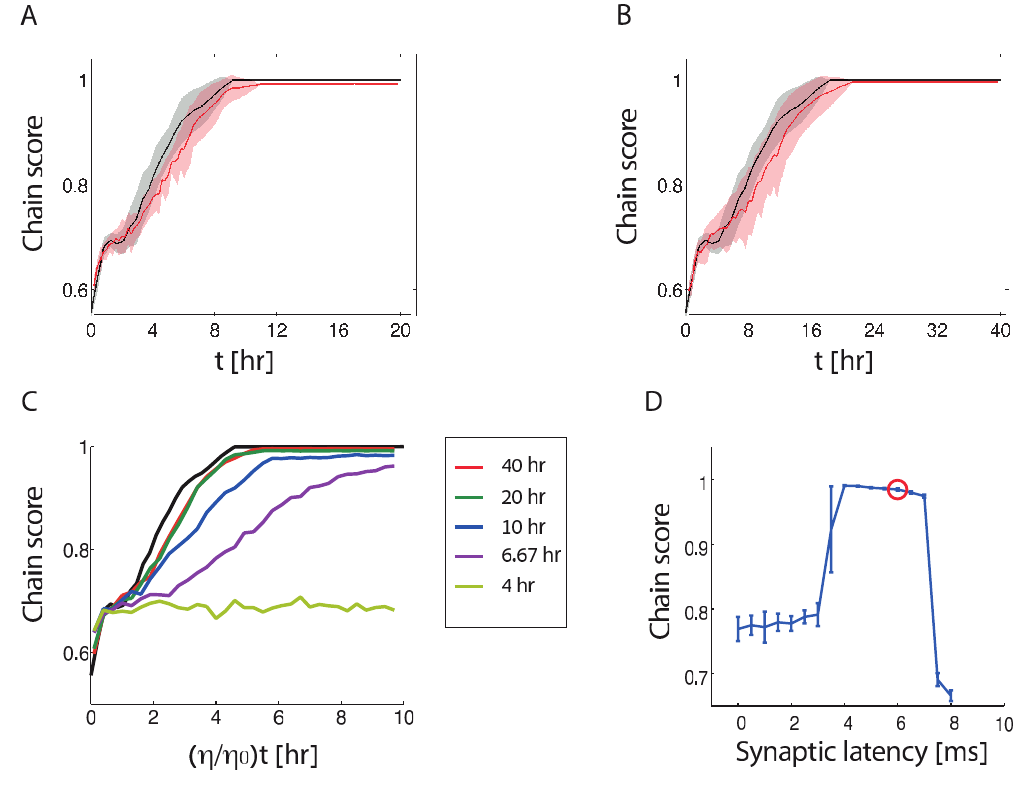}}
	\end{centerline}
	\vspace*{0.5cm}
	\protect\caption{\small \textbf{\label{fig:Synfire-stochastic}Formation of synfire chains under stochastic dynamics.} \textbf{A-B. }
		{Evolution in time of the chain score in the stochastic dynamics (red), compared with the prediction of the deterministic theory (black), shown for two different learning rates: $\eta = 2 \times 10^{-7}$ (A) and
			$\eta = 1\times 10^{-7}$ (B) (see \textit{Methods} for all other parameters). 
			Solid traces: average over ten initial conditions. Limits of the shaded areas represent the standard deviation. Note the different time scales in A and B. 
		}\textbf{C.}{ 
		Evolution in time of the chain score, shown for different learning rates ($\eta$) as a function of the time multiplied by $\eta/\eta_0$ where $\eta_0 = 4 \times 10^{-7}$.
		Each trace represents an average over ten random intial conditions, with a specific learning rate: $\eta=10^{-7}$ (red), $2 \times 10^{-7}$ (green),  $4\times 10^{-7}$ (blue), $6\times 10^{-7}$ (purple), and $10\times 10^{-7}$ (yellow). The black trace represents the prediction of the deterministic theory. 
		Note that the traces for the smallest learning rates (green and red traces, corresponding to panels A and B, respectively) nearly collapse on a single line, in close agreement with the prediction of the deterministic theory.
		The legend shows the actual time corresponding to the right end of the $x$ axis for each one of the plots. The synaptic latency $d = 6$\,ms. 
		\textbf{D}. Chain score evaluated in simulations of the stochastic dynamics, 
		shown as a function of the synaptic latency (the red circle corresponds to the synaptic latency in panel C). The learning rate $\eta = \eta_0 = 4 \times 10^{-7}$,
		matching the blue trace in panel C. Each data point
		represents an average over ten simulations with random initial conditions, and the error bars represent the standard deviation. The chain score is evaluated after 60 hours of simulated time. 
	}}
\end{figure}

These results demonstrate that the deterministic theory provides a relevant prediction for the stochastic synaptic drift when
the averaging is performed over a time scale of several minutes or more. 
However, during synfire chain formation the synaptic weights are constantly changing. 
Therefore, the whole process should occur over a significantly longer duration in order for the
deterministic theory to be adequate. In Fig~\ref{fig:Synfire-stochastic}A-C
we examine the plasticity dynamics in neural networks with initial
random synaptic weights, under parameters 
that enable robust emergence of synfire chain structures.
The chain score is measured in stochastic simulations with varying values of the learning rate $\eta$, 
and compared
with the prediction of the deterministic theory.
Each data point in the figure represents an average of the chain score over ten 
randomly chosen initial conditions. 

Within the deterministic theory the convergence time simply scales in proportion to $\eta^{-1}$.
In contrast, in the stochastic
simulations an increase in $\eta$ entails an increase in the amplitude of noise (relative to the
mean synaptic drift) and thus
we can expect the agreement with the deterministic theory to break down if $\eta$ is too large.
Figures \ref{fig:Synfire-stochastic}A,B demonstrate that the time course of the chain score in the 
stochastic simulations matches the deterministic theory very well when $\eta$ is set such that
convergence to synfire chains occurs over a time scale of $\sim$10 or $\sim$20 hours, respectively. 
Correspondingly, the time course of the chain score, 
when plotted as a function of $\eta t$
is nearly identical in these two conditions (red and green traces in 
Fig~\ref{fig:Synfire-stochastic}C), in very good agreement with the prediction of the deterministic 
theory (black trace). When the learning rate is faster, such that 
convergence to a synfire chain occurs over a time scale of $\sim$5 hours, 
the stochastic simulations are somewhat slower
to converge than the deterministic simulation (blue trace in 
Fig~\ref{fig:Synfire-stochastic}C). At an even higher learning rate, in which perfect
synfire chains emerge  
under the deterministic dynamics within $\sim$2 hours, 
 the stochastic simulations do not converge at all to synfire chain structures within a comparable time frame
 (yellow trace).

In summary, Figure \ref{fig:Synfire-stochastic} 
demonstrates that STDP in a stochastic spiking network can lead robustly
to synfire chain structures
over a time scale of several hours. Moreover, if the learning rate is sufficiently small, such that the 
time scale of convergence is
of order $\sim$10 hours or more, the time course of convergence is predicted very well by the deterministic theory of 
Eq.~\ref{eq: Presize STDP}. As another demonstration for the relevance of the deterministic theory, we measure 
in Fig~\ref{fig:Synfire-stochastic}D the
dependence of the chain score on the synaptic latency in stochastic spiking simulations lasting 60 hours of 
biological time, using the
same parameters as in Fig~\ref{fig:Synfire-chains _full_dynamics}C and a learning
rate that matches the blue trace in Fig~\ref{fig:Synfire-stochastic}C.
The results are very similar to those obtained in the deterministic dynamics 
(Fig~\ref{fig:Synfire-chains _full_dynamics}C).

\subsection*{Self organization into self connected assemblies}

\begin{figure}
\begin{centerline}
{\includegraphics[scale=0.7]{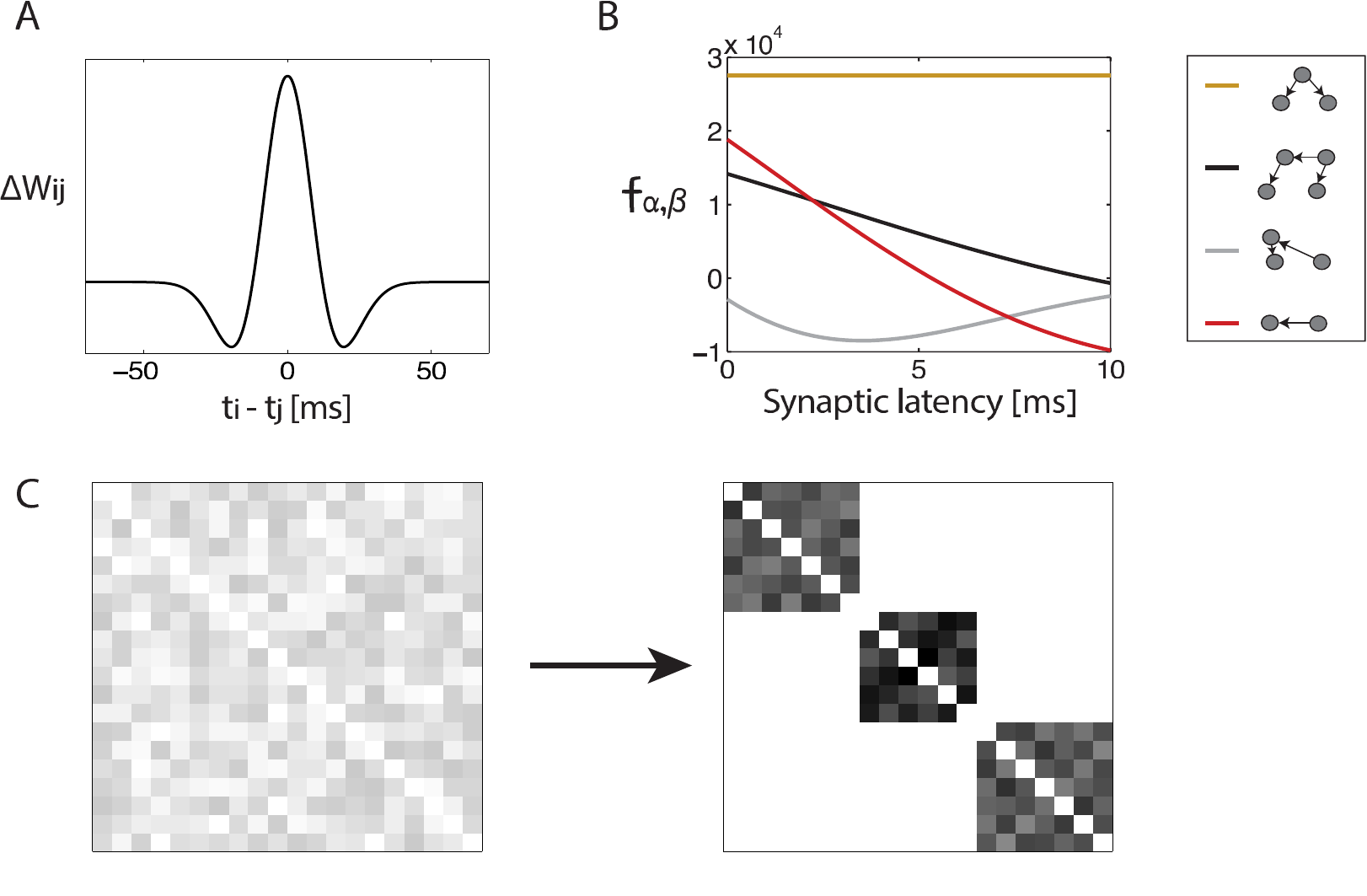}}
\end{centerline}
\vspace*{0.5cm}
\protect\caption{\small \textbf{\label{fig:assemblies}Self organization into
self connected assemblies.}{  }\textbf{A.}{
STDP function with Mexican hat structure.
}\textbf{B. }{The coefficients $f_{\alpha,\beta}$ of four different motifs as a function of the synaptic latency. Orange line - $\{1,1\}$, black line - $\{2,1\}$, gray line - $\{2,0\}$, red line - $\{1,0\}$.  
}\textbf{C.}{
Emergence of self connected assemblies under stochastic dynamics. Example of connectivity obtained in a parameter regime in which self connected assemblies emerge robustly. 
Left - initial connectivity, right - steady state.
Here the synaptic latency is $5.25$ ms (for other parameters, see \textit{Methods}).
}}
\end{figure}

Finally, we check whether self connected assemblies can  spontaneously emerge under the full STDP dynamics. 
The motif $\{1,1\}$ promotes formation of such structures (Fig~\ref{fig:complex structures}B,D). Therefore, we choose biophysical parameters that increase the relative contribution of this motif.
First, we choose an STDP function with a Mexican hat structure (Fig~\ref{fig:assemblies}A), which increases synaptic efficacies between neurons that spike at similar times, regardless of
the temporal order of the spikes. Second, we note that the contribution of the $\{1,1\}$ motif
is independent of synaptic latency, because both the pre and post synaptic 
neurons $i,j$ accrue the same latency relative to the source neuron $k$. On the other hand, coefficients of other low order motifs do depend on the synaptic latency (Fig~\ref{fig:assemblies}B). 

Based on the above reasoning, we expect that self connected assemblies will emerge, under the influence of STDP and synaptic competition, for finite synaptic latencies, in which 
contributions from motifs other than $\{1,1\}$ are suppressed. Figure~\ref{fig:assemblies}C shows results from a 
stochastic simulation of a
Poisson network, starting from initial random connectivity, with a synaptic delay of $\sim 5$\,ms. Here, a precise 
structure of self connected assemblies emerges robustly. Other structures were observed for alternative choices of the synaptic latency.

\section*{Discussion}
In summary, we developed a precise theoretical description of STDP in recurrent networks, which allows us to examine how the drift in the efficacy of each synapse depends on the full network structure. This dependence, although complicated, can be expressed as a sum of contributions from structural motifs, each with a fairly simple interpretation.  

High order motifs couple the plasticity dynamics of multiple synapses. Therefore, they can produce correlations between the synapses of different neurons, and promote the emergence of large-scale structures. Furthermore, under certain conditions, the high order motifs have a pivotal 
influence on the emerging connectivity. We demonstrated these central results
of the work within a scenario, in which STDP dynamics are combined with heterosynaptic competition. In this case, high order motifs
can drive the spontaneous formation of very ordered global structures,
such as wide synfire chains and self connected assemblies.
The same theoretical framework can be applied also to other scenarios,  in which STDP acts alone, or together with other plasticity mechanisms.

The analytical framework allows us to predict how the biophysical parameters, which characterize the dynamics
of single synapses, shape the emerging structures at the
level of the whole network. This relationship arises from the influence of the parameters on the relative contributions from different motifs. We varied the synaptic latency or the shape of the STDP function as an illustration of these dependencies, but other parameters such as the rise time of the synaptic current can have a similar influence on network plasticity.
It will also be interesting to consider in future 
work how the size of the network, and the desired size of clusters, influence the regime of parameters in which ordered structures are formed.

The influence of different structural motifs on spike correlation functions 
has been studied in several previous works \cite{Pernice2011,Trousdale2012,Hu2013}. Here, our interest
lies in the influence of these correlation functions on STDP acting on specific synapses within the network.
Thus, we view the spike 
correlations induced by each structural motif as the source for an effective interaction between the synapses
that participate in the motif.
For example, the motif of the form $\{1,1\}$, combined with a symmetric STDP function, 
encourages any two excitatory neurons that receive
presynaptic inputs from a common excitatory source to enhance the synaptic efficacies between them. 
The motif of the form $\{2,1\}$ encourages neurons that receive common input to project to common output units.

We worked with linear Poisson spiking neurons, since this approach allows us to analyze the consequences of STDP in a fully self-consistent manner, which does not involve any approximations beyond the initial choice of the underlying neural model (for another justification, see \cite{Pernice2012}). However, our analysis of STDP in terms of high order contributions from structural motifs is much more general. For example, Ocker et al. showed, very recently \cite{Ocker2015}, how STDP in recurrent neural networks of integrate and fire neurons can be analyzed based on an  expression for spike correlation functions, obtained under a linear response approximation. 
This approach leads to an expression for the synaptic drift which is very similar to the starting point of our analysis (Eq.~\ref{eq: Presize STDP}). Therefore, we expect that the qualitative consequences arising from high order motifs will be very similar under the two approaches.

\paragraph{High order motifs beyond the third order}
We focused on the influence of biophysical parameters on
the contribution of motifs up to third order. 
However, Eqs.~\ref{eq: Presize STDP} and \ref{eq:power decomposition} include also contributions from higher order motifs. This raises a question, why an analysis up to third order allowed us to predict the emergence of global structures. A partial answer to this question is that for small synaptic weights, the contribution of high order motifs decays with the number of participating synapses. One important factor contributing to the decay of motif coefficients is that when $\left|\alpha-\beta\right|$ is large the 
spike correlation function is shifted outside the range of the STDP window. In addition, spike correlation functions widen with increase of the 
number of participating synapses, thus decreasing their overlap with the STDP function. 

Another, more formal argument is based on the derivation of Eq.~\ref{eq:power decomposition}: as long as the synaptic weights are sufficiently weak, such that all eigenvalues
of the connectivity matrix are smaller than unity, 
the expansion in Eq.~\ref{eq: inverse expansion} converges for all $\omega$, and therefore the sum in Eq.~\ref{eq:power decomposition} 
must converge as well (this is also the condition for stability of the linear neural dynamics). 
This implies that the combined contribution from all motifs of order $n$ must decay as a function of $n$. 

Even though contributions of high order motifs must eventually decay, 
our simulations of the full STDP dynamics were performed in a regime where motifs beyond the third order do influence the plasticity.
To a large extent, the effect of higher order motifs can be predicted by the contribution of second and third order motifs (S5~Fig).
For example, all high order motifs that satisfy $\alpha-\beta=1$ are expected to assist the formation of synfire chains structures, based on the same intuition that was demonstrated in Fig~\ref{fig:complex structures}A. All these motifs also share a similar time course since they involve a delay of one synapse between the activity of the pre and post synaptic neurons. 
Similarly, all the motifs with $\alpha=\beta$ are expected to contribute to formation of self connected assemblies. In 
similarity to the motif $\{1,1\}$, and in contrast to the other motifs, their contribution is not influenced by the synaptic latency. The similar dependence
on the synaptic latency in motifs with the same value of $\alpha-\beta$ is illustrated in S5~Fig. 

\paragraph{Formation of synfire chains in previous works}
Fiete et al. \cite{Fiete2010} demonstrated that narrow synfire chains, in which single units project sequentially to each other, can emerge spontaneously under the 
combined influence of STDP and heterosynaptic competition. However, wide synfire chains did not emerge unless correlated inputs were fed into the network, even though the 
STDP simulations implicitly included motifs of all orders. Our results suggest why it was difficult to obtain wide synfire chains robustly in this work: 
first, Fiete et al. did not include in their model self depression, which can suppress the contribution of the first order motif, 
bringing the plasticity dynamics to an appropriate regime (see Fig \ref{fig:Synfire-chains motifs}A). In particular,
it is likely that the choice of parameters was such, that
the relative contribution of the third order motif was not sufficiently large.

Interestingly, wide (but sparsely connected) synfire chains were spontaneously
produced in another recent work \cite{Zheng2014}, which considered a 
 simplified
model of neural and synaptic dynamics, operating in discrete time bins.
By applying our framework to this model, it is 
 straightforward to see that
only a small subset
of the possible structural motifs contributed to plasticity, due to the simplified and discrete dynamics.
In addition, the synaptic plasticity rules included an effective form of self-depression. Thus, the
spontaneous formation of synfire chains in \cite{Zheng2014} is consistent with the predictions of our work. 

\paragraph{Distribution of synaptic weights}
Considerable theoretical attention has been devoted to the influence of STDP on the steady state distribution of synaptic weights
\cite{vanRossum2000,Rubin2001,Cateau2003,Gutig2003,Babadi2010,Luz2014,Effenberger2015}. 
This interest is
partially motivated by the 
observation in specific brain areas of unimodal distributions of synaptic efficacies, often 
following approximately a log-normal distribution 
\cite{Song2005,Ikegaya2013}.
Due to our interest in formation of synfire chains and
self-connected assemblies, we focused on situations in which the steady-state weight distribution is bimodal. However,
under certain choices of parameters which do not lead to the formation of ordered structures, we observe unimodal weight
distributions (see, for example, the black area in Fig~\ref{fig:Synfire-chains motifs}A 
which is characterized by strong synaptic self-inhibition 
\cite{Luz2014}). It will be of interest in future studies, to ask whether it is possible to obtain highly ordered structures,
in which the non-vanishing weights follow broad distributions, perhaps under a softer implementation of the synaptic 
competition.

It will also be interesting in future studies to consider situations in which connections exist  
between a subset of neuron pairs: for example, the structural connectivity may be sparse. 
The analytical framework that we developed can be directly applied to networks with an arbitrary adjacency matrix. In
this case, only efficacies of structurally existing synapses should be updated based on 
Eqs.~\ref{eq: Presize STDP}-\ref{eq:power decomposition} (note that in 
Eq.~\ref{eq:power decomposition}, only
those motifs that are realized in the structural connectivity graph can contribute to the sum, since the synaptic
efficacies associated with non-existing connections vanish).
Moreover, the formalism can be easily generalized to consider synapses with heterogeneous biophysical properties. 

\paragraph{Significance for neural dynamics and structure}
Nucleus HVC plays a key role in timing the vocal output of songbirds 
\cite{Hahnloser2002,Long2010}. This nucleus is a compelling candidate for a brain area that can organize autonomously to produce structured dynamics, since
auditory deprived songbirds generate a song with a stereotypical
temporal course \cite{Estes1965,Michael2002}. However, in almost all theoretical works that addressed how local plasticity rules give rise to temporal sequences of neural activity, it was necessary to provide some form of structured input into the network in order to robustly produce the sequential neural activity \cite{Suri2002,Fiete2010,Klampfl2013,Bayati2015}. Similarly, structured inputs were required in order to robustly produce self connected assemblies, which give rise to another useful form of neural dynamics, characterized by multiple stable states
\cite{Amit1997,Wang2002,Renart2007,Litwin-Kumar2012,Stern2014}. 
It is therefore significant that synaptic structures which support structured neural dynamics 
can emerge in a neural circuit without any preexisting order in the synaptic organization, and without any exposure to external stimuli.
Appropriate choices of the biophysical parameters, which enable this type of autonomous organization, became apparent by applying the theoretical formalism and reasoning developed in this work.

Finally, we briefly mention another area of future work, in which the formalism developed here may find a useful application: we expect that the analysis of synaptic dynamics in terms of contributions from structural motifs, will be valuable for assessing the role of STDP in shaping the high order statistics of cortical connectivity, as experimental data on these statistics become increasingly available \cite{Song2005,Yoshimura2005,berger2011}.

\section*{Methods}

\subsection*{Network dynamics }

The time dependent activity of neuron $i$  is a stochastic realization of an inhomogeneous Poisson process, with expectation value

\begin{equation}
\lambda_{i}\left(t\right)= \sum_{k=1}^{N}W_{ik}\int_{-\infty}^{t}{\rm d} t'a\left(t-t'\right) S_{k}\left(t'\right)+b_{i} \,,
\label{eq:network dynamics}
\end{equation}
where $N$ is the number of neurons,  $W$ is the connectivity matrix, $a\left(t\right)$
is the synaptic current, $b_{i}$ is a constant external input, and $S_{k}\left(t\right)=\sum_{\mu}\delta\left(t-t_k^{\mu}\right)$
is the spike train of the neuron $k$ (where $t_k^{\mu}$ are spike times of the  neuron).
We assume that the neurons do not excite themselves, meaning
that $\forall i\mbox{ }W_{ii}=0$. 

\subsection*{An analytical expression for the synaptic drift}

Assuming that all spike pairs contribute to STDP, the change in the synaptic efficacies due to STDP can be expressed as follows:
\begin{equation}
\Dot{W}_{ij}^{{\rm STDP}}\left(t\right)=\dot{W}_{ij}^{+}\left(t\right)+\dot{W}_{ij}^{-}\left(t\right),
\label{eq:explicit_STDP term}
\end{equation}
where 
\begin{equation}
\dot{W}_{ij}^{+}\left(t\right)=S_{i}\left(t\right)\int_{-\infty}^{t}S_{j}\left(t'\right)F\left(t-t'\right){\rm d}t'
\label{eq:wdotplus}
\end{equation}
is the change in synaptic efficacy arising from spikes in the post synaptic neuron $i$ at time $t$, and
\begin{equation}
\dot{W}_{ij}^{-}\left(t\right)=S_{j}\left(t\right)\int_{-\infty}^{t}S_{i}\left(t'\right)F\left(t'-t\right){\rm d}t'
\label{eq:wdotminus}
\end{equation}
is the change following a spike in the pre synaptic neuron $j$ at time $t$. In both terms, the integration is over all previous spikes of the presynaptic neuron (Eq.~\ref{eq:wdotplus}) or the postsynaptic neuron (Eq.~\ref{eq:wdotminus}). 

\begin{figure}[h]
	\begin{centerline}
		{\includegraphics[scale=0.58]{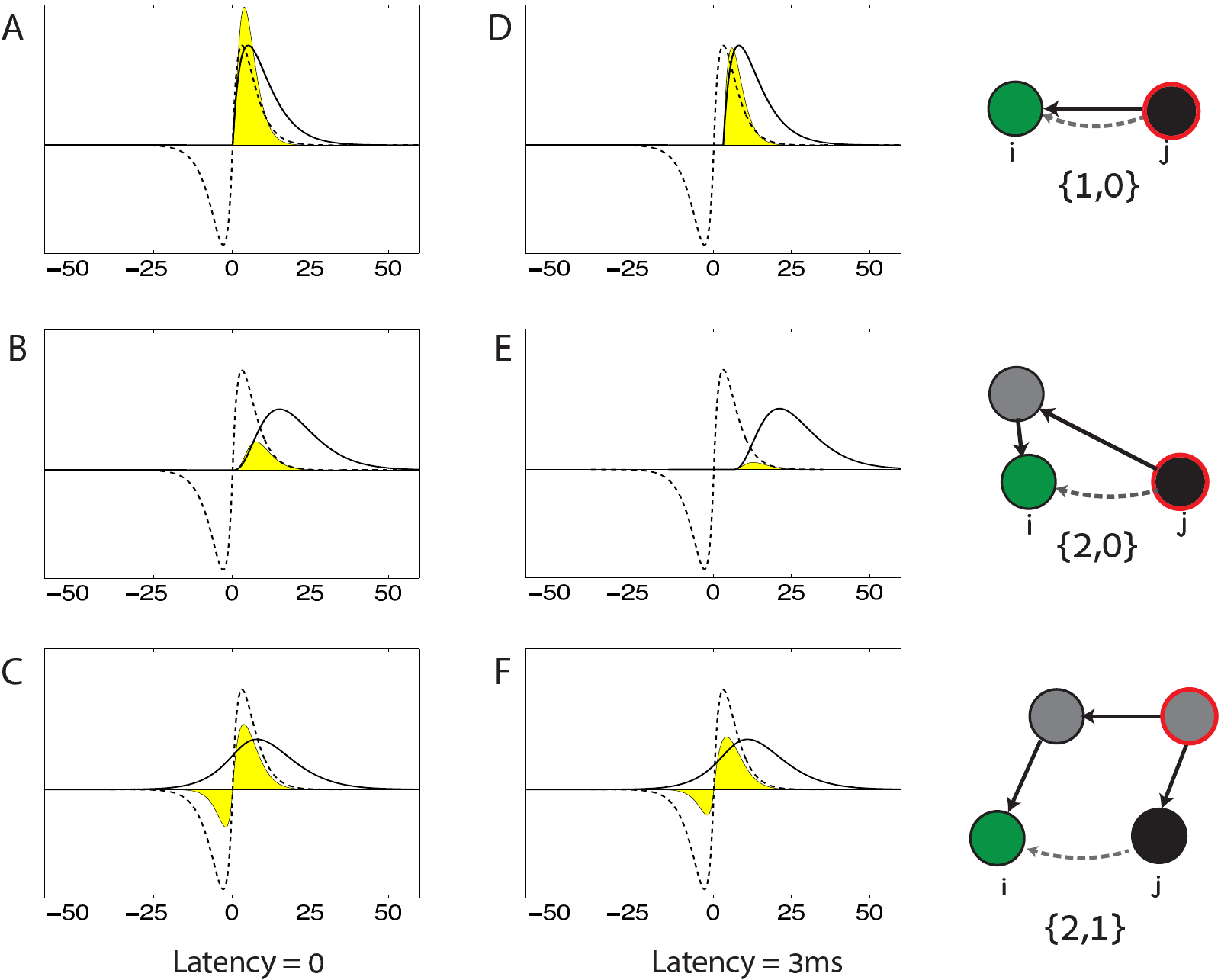}}
	\end{centerline}
	\protect\caption{\footnotesize  \textbf{\label{motifs methods} Dependence of motif coefficients on the time course of STDP and time course of post synaptic currents.} 
		\textbf{A-F}. Examples illustrating how the coefficients $f_{\alpha, \beta}$ are determined for three motifs.
		Solid lines: motif driven spike correlations (Eq.~\ref{eq:c_alpha_beta}). Dashed lines: 
		the STDP function (chosen to be antisymmetric). Yellow area: the integral of the product
		of these two functions, which determines the motif coefficient. 
		For simplicity, we consider only the correlation structure induced by the excitatory synaptic connectivity.
		\textbf{A}. The correlation time course (solid line) induced by the  motif $\{1,0\}$ is simply the synaptic current function. 
		\textbf{B}. In comparison with A,
		the correlation time course induced by the motif $\{2,0\}$ peaks later and is more widespread, 
		because the coupling between the pre and the post synaptic neurons is mediated through an intermediate neuron. 
		\textbf{C}. The correlation function induced by the third order motif $\{2,1\}$ peaks at a similar time as in A, 
		because in both motifs $\alpha - \beta = 1$, meaning that the paths leading to neurons $i$ and $j$ differ in length by one synapse. 
		\textbf{D-F}. A synaptic latency of 3\,ms affects the overlap between the correlation time course and the STDP function, and thus the
		motif coefficients (see also Fig~\ref{fig:-Motifs coefficients}). 
		\textbf{D}. In the motif $\{1,0\}$ the correlation time course is shifted by 3\,ms relative to panel A, leading to a decrease in the overlap with the STDP function
		and a decrease in $f_{1,0}$. 
		\textbf{E}. In the motif $\{2,0\}$ the spike correlation function (solid line)
		is shifted by 6\,ms relative to panel B because two synaptic latencies 
		accumulate along the path from the pre synaptic neuron $j$ to the post synaptic neuron $i$. The overlap with the STDP function and the coefficient $f_{2,0}$
		are sharply reduced.
		\textbf{F}. In the motif $\{2,1\}$ the spike correlation function is shifted by 3\,ms relative to panel because the relative latency, 
		incurred along the paths from the source neuron to neurons $i$ and $j$ amounts to a single synaptic latency.
		The time course of correlations arising from this motif (solid line) is very broad, and therefore shifting it relative
		to the STDP trace has relatively little influence on the overlap with the STDP function and on $f_{2,1}$.
	}
\end{figure}

We define the correlation function of spikes in each pair of cells as follows,
\begin{equation}
C_{ij}\left(\tau\right)\equiv\left\langle S_{i}\left(t+\tau\right)S_{j}\left(t\right)\right\rangle\,, 
\label{eq:correlations definition}
\end{equation}
where $\left\langle \cdot\right\rangle $ denotes averaging over different realizations of the Poisson dynamics for a given connectivity. For constant external input, and under the assumption of slow learning, the correlation function is stationary, and does not depend on $t$. 
We denote the rate of change in the synaptic efficacy, averaged over realizations of the Poisson dynamics as
\begin{equation}
\Delta_{ij}^{{\rm STDP}} \equiv
\left\langle \dot{W}_{ij}^{\rm STDP} \right\rangle\,,
\end{equation}
and refer to it in short as the synaptic drift. 
Using the correlation function we can express the synaptic drift as follows,
\begin{equation}
\Delta_{ij}^{{\rm STDP}}
=\int_{-\infty}^{\infty}C_{ij}\left(\tau\right)F\left(\tau\right)d\tau\,.
\label{eq:STDP drift}
\end{equation}

For linear dynamics the correlation function can be written exactly. In the frequency domain
\cite{The1971},
\begin{equation}
\tilde{C}\left(\omega\right)=
2\pi \delta\left(\omega\right) rr^{T}
+\left[I-\tilde{a}\left(\omega\right)W\right]^{-1}D\left[I-\tilde{a}\left(-\omega\right)W^{T}\right]^{-1}\,,
\label{eq:Corelation Hawkes}
\end{equation}
where $r_{i}=\left\langle \lambda_{i}\right\rangle $, the diagonal matrix $D_{ij}=\delta_{ij}r_{j}$,  $I$ denotes the unit matrix, and we use the convention in which the Fourier transform of a function $g(t)$ is defined as
$\tilde{g}(\omega) = \int_{-\infty}^{\infty} {\rm e}^{-i \omega t}g(t){\rm d}t$.

The average firing rate can be easily obtained from Eq.~\ref{eq:network dynamics}:
\begin{equation}
r =\left[I-\tilde{a}\left(0\right)W \right]^{-1} b\,.
\label{eq:average firing rate calculation}
\end{equation}
By substituting Eq.~\ref{eq:Corelation Hawkes} in 
Eq.~\ref{eq:STDP drift}, we obtain Eq.~\ref{eq: Presize STDP}. 
Note that the diagonal terms should be ignored, since we assume that there is no synapse
from a neuron to itself. 

\paragraph{Power series expansion of the average learning dynamics}

The matrix inverses appearing in Eq.~\ref{eq:Corelation Hawkes} can be expanded using the matrix identity \cite{Pernice2011}
\begin{equation}
\left[I-A\right]^{-1}=\sum_{i=0}^{\infty}A^{i}\,.
\label{eq: inverse expansion}
\end{equation}
In our context, this substitution can be seen as an expansion in powers
of the synaptic efficacies. We assume that the synaptic efficacies are sufficiently weak, such that for all values of $\omega$, all eigenvalues of
$\tilde{a}\left(\omega\right)W$ are smaller (in absolute magnitude) than unity.
Note that below we normalize the synaptic current such that $\tilde{a}(0) = 1$,
and $\left|\tilde{a}(\omega)\right| \leq \left|\tilde{a}(0)\right|$ for all $\omega$. In this case the requirement is that all eigenvalues of $W$ are smaller
(in absolute magnitude) than unity.

Using  Eq.~\ref{eq: inverse expansion} we rewrite Eq.~\ref{eq: Presize STDP} as Eq.~\ref{eq:power decomposition}, where we define:
\begin{equation}
f_{\alpha,\beta} 
=\frac{1}{2\pi}\int_{-\infty}^{\infty}{\rm d}\omega\tilde{F}\left(-\omega\right)
\tilde{a}\left(\omega\right)^{\alpha}\tilde{a}\left(-\omega\right)^{\beta}
\label{eq:motif coefficients}
\end{equation}
and
\begin{equation}
f_{0} 
=\tilde{F}\left(0\right)\,.
\label{eq:f_0_coefficient}
\end{equation}
Note that $f_{\alpha,\beta}$ are dimensionless, and $f_0$ has dimensions of time.
In the time domain, 
\begin{equation}
f_{\alpha, \beta} = \int_{-\infty}^{\infty}{\rm d}t \, F(t) \cdot c_{\alpha, \beta}(t) 
\end{equation}
where $c_{\alpha, \beta}$ can be written as a series
convolutions of the synaptic current function:
\begin{equation}
c_{\alpha, \beta}(t) = \underbrace{a(t) * \ldots * a(t)}_{\alpha \text{ terms}} * \underbrace{a(-t) * \ldots * a(-t)}_{\beta \text{ terms}}\,.
\label{eq:c_alpha_beta}
\end{equation}
The way $f_{\alpha, \beta}$ is determined from the synaptic current and the STDP function
is illustrated in Fig~\ref{motifs methods} for a few examples.

\subsection*{Full plasticity dynamics}

We consider N excitatory neurons, with modifiable recurrent connections. 
In addition to the STDP, the excitatory synapses undergo heterosynaptic competition, and possibly constant self depression and
growth at a constant rate.
The full plasticity dynamics of these synapses can be summarized by the following expression:

\begin{equation}
\dot{W}_{ij}^{{\rm ex}}\left(t\right)=
\eta\left[\dot{W}_{ij}^{{\rm STDP}}\left(t\right)-\psi\Delta_{i}^{{\rm in}}-\psi\Delta_{j}^{{\rm out}}-\mu W_{ij}^{\rm ex}+\gamma\right] \,.
\label{eq:Complete learning rule}
\end{equation}
The terms $\Delta_{i}^{{\rm in}}\mbox{, }\Delta_{j}^{{\rm out}}$ represent the heterosynaptic competition \cite{Fiete2010}: 
competition over the input to neuron $i$,
\begin{equation}
\Delta_{i}^{{\rm in}}=\left(\sum_{k}W_{ik}^{{\rm ex}}-W_{\rm max}\right)\cdot\Theta\left[\sum_{k}W_{ik}^{{\rm ex}}-W_{\rm max}\right]\,,
\end{equation}
and competition over the outputs from neuron $j$:
\begin{equation}
\Delta_{j}^{{\rm out}}=\left(\sum_{k}W_{kj}^{{\rm ex}}-W_{\rm max}\right)\cdot\Theta\left[\sum_{k}W_{kj}^{{\rm ex}}-W_{\rm max}\right]\,.
\end{equation}
Here, $\Theta\left(x\right)$ is the Heaviside step function: 

\begin{eqnarray*}
\Theta\left(x\right) & = & \begin{cases}
0 & x<0\\
1 & x\geqq0
\end{cases}
\end{eqnarray*}
When $\psi$ is sufficiently large, the competition guarantees that the sum over each row and column of $W^{\rm ex}$ does not
exceed $W_{\rm max}$. 
Finally, the term $\mu W_{ij}^{\rm ex}$ represents self depression (constant weakening of the synapse in proportion to the synaptic efficacy), 
and the term $\gamma$ represents a constant growth of each weight at a fixed rate. 

In addition to these rules,
the excitatory synapses are restricted to the range $[0,w_{{\rm max}}]$.

\paragraph{Inhibitory synapses}

In all plasticity simulations, the  connectivity between each pair of neurons was divided into excitatory and inhibitory components,

\begin{equation}
W=W^{{\rm ex}}+W^{\rm in}, 
\end{equation}
where $W^{\rm ex}$ are the excitatory synaptic efficacies, which satisfy the dynamics described above, and $W^{\rm in}$ are effective inhibitory synaptic efficacies. These have the following structure:
\begin{equation}
W^{\rm in}_{ik}=\frac{1}{N}
\sum_{l}W^{\rm ex}_{il}
\,. 
\end{equation}
Because $W_{ik}$ does not depend on $k$, this form of inhibition can 
be interpreted as arising from an inhibitory drive which is proportional
to the total activity within the network, and is mediated by inhibitory interneurons with fast synapses. 
The sum of inhibitory synapses into each neuron is dynamically adjusted to balance the sum of the excitatory synapses.

\subsection*{Simulations}

\paragraph{Power series expansion simulations}

In Fig~\ref{fig:complex structures},\ref{fig:Synfire-chains motifs}, Fig~\ref{fig:Synfire-chains _full_dynamics}B,C (Gray and red lines), and 
in S2~Fig,
STDP is modeled as in
Eq.~\ref{eq:power decomposition}, but with a small number of non-vanishing coefficients $f_{\alpha, \beta}$ which are set as in Table \ref{table:f}.
In Fig~\ref{fig:Synfire-chains _full_dynamics}B,C the coefficients $f_{\alpha, \beta}$ which are not set to zero
are directly evaluated from the STDP function and the synaptic current function.

\begin{table}
\centering
\FloatBarrier
    \begin{tabular}{ | c | c | c |c |c | c |c |c |}
    \hline
    \rowcolor{lightgray}
      & $f_{1,0}$ & $f_{0,1}$ & $f_{2,0}$ & $f_{0,2}$ & $f_{1,1}$ & $f_{2,1}$ & $f_{1,2}$\\
      \hline
    Fig~\ref{fig:complex structures}C & $1$ & $-1$ & $1$  & $-1$ & $0$ &\begin{tabular}[x]{@{}c@{}}left panel: 0\\right panel: $16$\end{tabular} & $-f_{2,1}$ \\ [2ex]
    \hline
    Fig~\ref{fig:complex structures}D & $0.25$ & $0.25$ & $0$ &$0$ &\begin{tabular}[x]{@{}c@{}}left panel: 0\\right panel: $1.8$\end{tabular} & $0$ & $0$\\
    \hline
   Fig~\ref{fig:Synfire-chains motifs}A  & * & * & $0$ & $0$ & $0$ & $20$ & $-20$\\
     \hline
     Fig~\ref{fig:Synfire-chains motifs}B & * & $-30$ & $0$ & $0$ &$0$ & * & $-f_{2,1}$\\[2ex]
     \hline
      S2~Fig~A & $-2$ & $-30$ & * & $-f_{2,0}$ & $0$ & * &$-f_{2,1}$\\[2ex]
    \hline
    S2~Fig~B  & $-2$ & $-30$ & $0$ & $0$ & * & * &$-f_{2,1}$ \\[2ex]
    \hline
    S2~Fig~C  & * & * & $0$ & $0$ & $0$ & $20$ & $-20$\\
     \hline
    \end{tabular}
    \\
     * Values are specified in the Figure.
    \caption{The coefficients $f_{\alpha,\beta}$ in Figs \ref{fig:complex structures},\ref{fig:Synfire-chains motifs} and S2~Fig.}
    \label{table:f}
     \end{table}

 \paragraph{Average learning simulations}

In the average learning simulations (Figs~\ref{fig:Synfire-chains _full_dynamics},\ref{fig:Synfire-stochastic}, S1, S3, and S4), the 
contribution of STDP to the synaptic drift was evaluated numerically in the Fourier domain,
using Eq.~\ref{eq: Presize STDP}. 

\paragraph{Stochastic Poisson learning simulations}

In the stochastic simulations (Figs~\ref{fig:Synfire-stochastic},\ref{fig:assemblies}, and S1~Fig) the STDP term was explicitly determined
by the spiking activity in the network (Eqs.~\ref{eq:explicit_STDP term}-\ref{eq:wdotminus}).

\paragraph{Simulations parameters}

\textit{Synaptic current function}:
In all simulations,
\begin{equation}
a\left(t\right)=\begin{cases}
a_{0} \,{\rm exp}\left(\displaystyle -\frac{t-d}{\tau_{1}}\right)
\left[1-{\rm exp}\left(\displaystyle -\frac{t-d}{\tau_{2}}\right)\right] & \ \ \  t>d\\
0 & \ \ \ t<d
\end{cases}\, ,
\end{equation}
where $\tau_{1}=5{\rm ms}$, $\tau_{2}=1\,{\rm s}$, and $d$
is the synaptic latency. The coefficient $a_{0}$ normalizes the synaptic current such that $\int_{-\infty}^{\infty}a\left(t\right)=1$. This choice sets a meaningful
scale for $W$: if $W_{ij} = 1$, a single spike in a pre synaptic neuron $j$ increases (or decreases), on average, the 
number of spikes emitted by the post synaptic neuron $i$ by one. 

\medskip

\noindent \textit{STDP function}:
In Figs \ref{fig:-Motifs coefficients}, \ref{fig:Synfire-chains _full_dynamics},  \ref{fig:Synfire-stochastic}, 
\ref{motifs methods}, S1 and S3
we used the following antisymmetric STDP function:
\begin{equation}
F\left(t\right)= h_{0}\cdot h\left(t\right)\,,
\end{equation}
where:
\begin{equation}
h\left(t\right)=\begin{cases}
A_{+}\,{\rm exp}\left(-\frac{\displaystyle t}{\displaystyle \tau_{1}}\right)
\left[1-{\rm exp}\left(-\frac{\displaystyle t}{\displaystyle \tau_2}\right)\right] 
& \ \ \ t>0\\
A_{-}\,{\rm exp}\left(\frac{\displaystyle t}{\displaystyle \tau_{1}}\right)
\left[1-{\rm exp}\left(\frac{\displaystyle t}{\displaystyle \tau_2}\right)\right]
& \ \ \ t<0
\end{cases}\,,
\end{equation}
$h_{0}=10^4$, $\tau_{1}=3{\rm ms}$, $\tau_{2}=2{\rm sec}$, and
$A_{+}=\frac{\displaystyle 0.8}{\displaystyle \tau_{1}}$, $A_{-}=-A_{+}$.

In Fig~\ref{fig:assemblies} we used a symmetric STDP function
\begin{equation}
F\left(t\right)=A \left[1-\frac{t^{2}}{\sigma^{2}}\rm{exp}\left(-\frac{8t^{2}}{5\sigma^{2}}\right)\right]\,,
\end{equation}
where $\sigma=12$ms, $A=5.2\cdot10^4$.

In S4~Fig and S5~Fig  we used the following asymmetric (but not antisymmetric) STDP function:

\begin{equation}
F\left(t\right)= h_{0}\cdot h\left(t\right)\,,
\end{equation}
where:
\begin{equation}
h\left(t\right)=\begin{cases}
A_{+}\,{\rm exp}\left(-\frac{\displaystyle t}{\displaystyle \tau_{1_+}}\right)
\left[1-{\rm exp}\left(-\frac{\displaystyle t}{\displaystyle \tau_2}\right)\right] 
& \ \ \ t>0\\
A_{-}\,{\rm exp}\left(\frac{\displaystyle t}{\displaystyle \tau_{1_-}}\right)
\left[1-{\rm exp}\left(\frac{\displaystyle t}{\displaystyle \tau_2}\right)\right]
& \ \ \ t<0
\end{cases}\,,
\end{equation}
$h_{0}=150\int_{\infty}^{\infty}h\left(t\right)dt$, $\tau_{1_{+}}=3{\rm ms}$, $\tau_{1_{-}}=4.5{\rm ms}$, $\tau_{2}=2{\rm sec}$,
$A_{+}=0.8/\tau_{1_+}$, and $A_{-}=0.525/\tau_{1_-}$. 

\medskip

\noindent \textit{Limitation on the total synaptic input
and output}:
In all simulations,
\begin{equation}
\label{eq: M}
W_{\rm max}=M\cdot w_{\rm max}\,.
\end{equation}
This parameter affects the number of neurons in each group when the ordered structures emerge \cite{Fiete2010}. 

\medskip

\noindent \textit{Learning rate}:
In all simulations except the stochastic simulations, the learning dynamics were 
implemented using the Euler method with an adaptive time step, chosen such that the maximal
change in the weights in each step was $0.0005$ (Fig~\ref{fig:complex structures}), $0.0001$ (Fig~\ref{fig:Synfire-chains motifs} and S2~Fig) and $0.02$ (Fig~\ref{fig:Synfire-chains _full_dynamics} , S3 and S4), and $0.002$ (Fig~\ref{fig:Synfire-stochastic}).

\medskip

\noindent \textit{Initial connectivity}:
The initial weights were chosen independently from a uniform distribution between 0 and $w_{\rm max}M/N$ (Fig~\ref{fig:complex structures},\ref{fig:Synfire-chains motifs},\ref{fig:assemblies} and S2~Fig), and between 0 and  $1.5w_{\rm max}M/N$ (Fig~\ref{fig:Synfire-chains _full_dynamics},\ref{fig:Synfire-stochastic}, S3 and S4). 

\medskip

\noindent \textit{Simulations duration and convergence criterion}:
The convergence criterion was that in the last 10 iterations the absolute
change in each element of the matrix did not exceed $e^{-15}$. 
The stochastic simulations were conducted using stochastic Euler dynamics with a time step 
of $0.25$\,ms.

\medskip

\noindent Other parameters are summarized in Tables~\ref{parameters},\ref{parameters_suplementary}.

\begin{table}
\small
\centering
    \begin{tabular}{ | l | l | l |l |l | l|}
    \hline
    \rowcolor{lightgray}
      & Fig~\ref{fig:complex structures} & Fig~\ref{fig:Synfire-chains motifs}&Fig`\ref{fig:Synfire-chains _full_dynamics} & Fig~\ref{fig:Synfire-stochastic} & Fig~\ref{fig:assemblies}\\
       \hline
    $N$ & \begin{tabular}[x]{@{}c@{}}Fig~\ref{fig:complex structures}C: 20\\Fig~\ref{fig:complex structures}D: 24\end{tabular} & 20
    &20 & 20 & 20  \\[2ex]
      \hline
    $\psi$ &
    $5\cdot10^{4}$\,s$^{-1}$  & 
    $2\cdot10^{3}$\,s$^{-1}$ &
    $5\cdot10^{4}$\,s$^{-1}$ &
    $5\cdot10^{4}$\,s$^{-1}$ & 
    $ 10^{5}$\,s$^{-1}$   \\ [2ex]
    \hline
    $\eta$ & $10^{-8}$&$10^{-8}$ & $10^{-8}$ & \begin{tabular}[x]{@{}c@{}}
    Fig~\ref{fig:Synfire-stochastic}A: $2\cdot10^{-7}$\\
    Fig~\ref{fig:Synfire-stochastic}B: $1\cdot10^{-7}$\\
    Fig~\ref{fig:Synfire-stochastic}C: *\\
    Fig~\ref{fig:Synfire-stochastic}D: $4\cdot10^{-7}$\end{tabular} 
    &$2.5\cdot10^{-9}$  \\[1.5ex]
    \hline
    $b$ & 0 & 15Hz & 15Hz & 15Hz & 5Hz \\
     \hline
    $\mu$ & -- &--&\begin{tabular}[x]{@{}c@{}}
    Fig~\ref{fig:Synfire-chains _full_dynamics}B: *\\
    Fig~\ref{fig:Synfire-chains _full_dynamics}C: $300b$\end{tabular}  & $300b$ & 0 \\[2ex]
     \hline
    $M$ & 5 &5 &5  & 5  & 4   \\ 
   \hline
    $w_{\rm max}$ &$\displaystyle{\frac{0.85}{M}}$ &$\displaystyle{\frac{0.85}{M}}$ & $\displaystyle{\frac{0.9}{M}}$ & $\displaystyle{\frac{0.9}{M}}$ &$\displaystyle{\frac{0.9}{M}}$ \\[2ex]
    \hline
    $\gamma$ & $0$ & $225$s$^{-1}$ & $225$s$^{-1}$ &$225$s$^{-1}$ &$0$\\[2ex]
   \hline
    $d$ &- & - & \begin{tabular}[x]{@{}c@{}}
    Fig~\ref{fig:Synfire-chains _full_dynamics}B: $0$\\
    Fig~\ref{fig:Synfire-chains _full_dynamics}C: *\end{tabular} 
 & \begin{tabular}[x]{@{}c@{}}
    Fig~\ref{fig:Synfire-stochastic}A: $6$ ms\\
    Fig~\ref{fig:Synfire-stochastic}B: $6$ ms\\
    Fig~\ref{fig:Synfire-stochastic}C: $6$ ms\\
    Fig~\ref{fig:Synfire-stochastic}D: * \end{tabular} 
 &$5.25$ ms\\[2ex]
    \hline

    \end{tabular}
   \\
   * Values are specified in the Figure.
   \caption{Simulation parameters}
       \label{parameters}
    \end{table}

\begin{table}
\small
\centering
    \begin{tabular}{ | l | l | l |l |l |}
    \hline
    \rowcolor{lightgray}
      & S1 Fig & S2 Fig & S3 Fig & S4 Fig\\
       \hline
    $N$ & 20 & 20
    &20 & 20   \\[2ex]
      \hline
    $\psi$ &
    --  & $2\cdot10^{3}$\,s$^{-1}$
    &
    $5\cdot10^{4}$\,s$^{-1}$ &
    $5\cdot10^{4}$\,s$^{-1}$ 
    \\ [2ex]
    \hline
    $\eta$ & $4\cdot10^{-7}$& $10^{-8}$ & $10^{-8}$ & $10^{-8}$  \\[1.5ex]
    \hline
    $b$ &  15Hz & 15Hz & 15Hz & 15Hz  \\
     \hline
    $\mu$ & -- &--&$300b$ & $600b$  \\[2ex]
     \hline
    $M$ & 5 &5 &4  & 4     \\ 
   \hline
    $w_{\rm max}$ &$\displaystyle{\frac{0.9}{M}}$ &$\displaystyle{\frac{0.85}{M}}$ & * & $\displaystyle{\frac{0.9}{M}}$ 
    \\[2ex]
    \hline
    $\gamma$ & $225$s$^{-1}$ & $225$s$^{-1}$ & $225$s$^{-1}$ &$0$ \\[2ex]
    \hline
 $d$ & $0$ &- & * & * \\[2ex]
    \hline
    \end{tabular}
   \\
   * Values are specified in the Figure.
   \caption{supplementary parameters}
       \label{parameters_suplementary}
    \end{table}

\subsection*{Chain score and ordering of connectivity matrices}

To classify groups of neurons that share similar connectivity,
we performed k-means classification
on a set of $N$ vectors, where the $i$-th vector includes all the excitatory input and output synaptic efficacies of neuron $i$: $\{W_{ik}^{{\rm ex}}\mbox{,    }W_{ik}^{^{T}{\rm ex}}\}_{k=1\cdots N}$,
and using a squared Euclidean distance. 
We then reordered the neurons (and the connectivity matrix) based on the groups identified
by the k-means clustering. When searching for synfire chain structure, we chose the order of groups 
as follows: We randomly chose one group and set it as the first group. We then looked for a remaining group that receives the largest total synaptic input from the first group, and set it as the next group. This process was repeated to include all groups. 
Next, we compared the ordered connectivity matrix to an
``ideal'' binary connectivity matrix that represents complete feed
forward connectivity between the groups, or complete clustering into self connected groups.
The chain score is defined as: $1-x$, where $x$ is the normalized square distance
between the ordered matrix, scaled by the largest element, and a matrix representing
ideal feed forward connectivity (or a perfect arrangement of self connected clusters).
Finally we maximized the similarity score over a range of values of $k$. 

\section*{Acknowledgments}
We thank Mor Nitzan and Idan Segev for useful discussions, and
Nimrod Shaham for comments on the manuscript.


\pagebreak
\section*{Supporting Information}
\renewcommand{\thefigure}{S\arabic{figure}}
\setcounter{figure}{0}

\begin{figure}[b]
\begin{centerline}
{\includegraphics[scale=0.9]{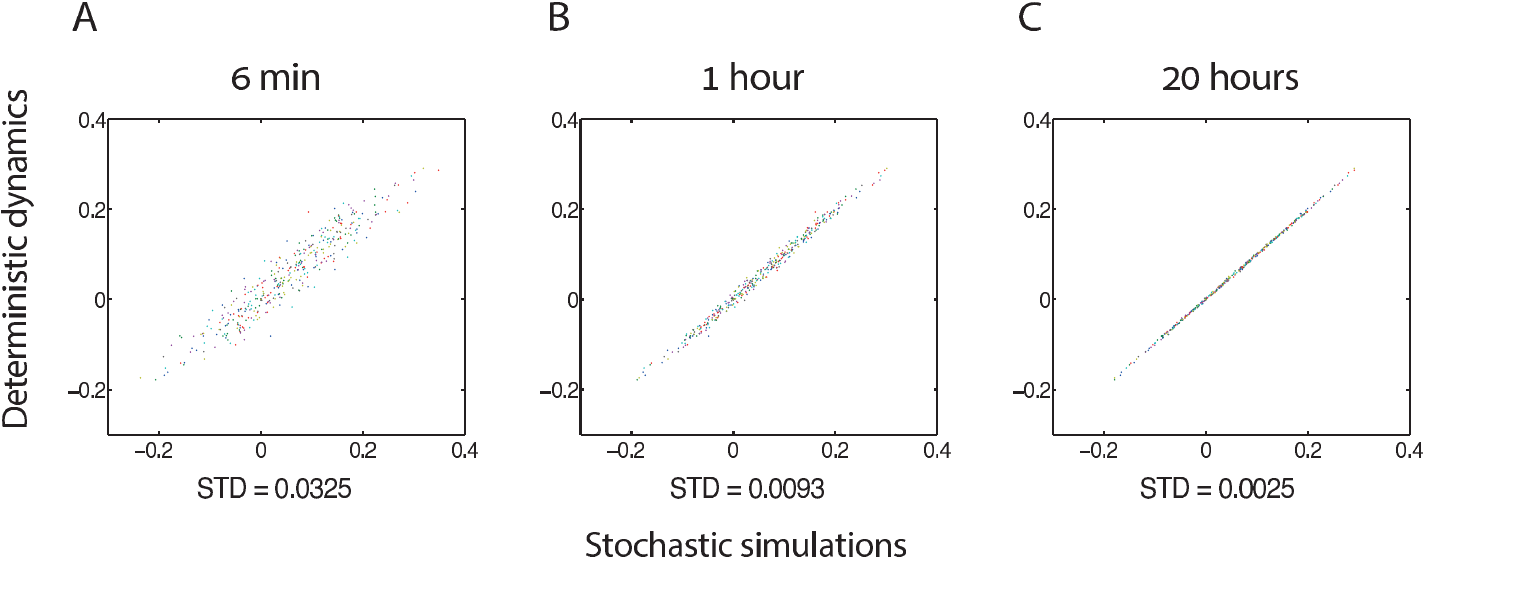}}
\end{centerline}
\protect\caption{\small \textbf{\label{stochastic vs average} Comparison between the deterministic dynamics and the stochastic dynamics. } 
Scatter plot of the analytical expression for the average change in the synaptic efficacy (vertical axis), against the change in the synaptic efficacy generated by STDP 
(Eqs.~\ref{eq:explicit_STDP term}-\ref{eq:wdotminus})
in a stochastic simulation of
stochastic spiking neurons, averaged
over 
6 minutes (\textbf{A}), 1 hour (\textbf{B}), and 20 hours (\textbf{C}). Each point represents the change in one synapse, where all synapses belong to the same connectivity matrix. 
The standard deviation (STD) listed under each panel is the mean square distance, across all synaptic pairs, between the change of the synaptic efficacy in the stochastic simulation
and the prediction of the deterministic theory.
The synaptic efficacies are drawn independently from a uniform distribution between $\left[0,\frac{2}{N}W_{\rm max}\right]$, where $W_{\rm max}=0.9$, $N=20$,
and are kept fixed during the stochastic simulations.}
\end{figure}

\begin{figure}[b]
\begin{centerline}
{\includegraphics[scale=0.8]{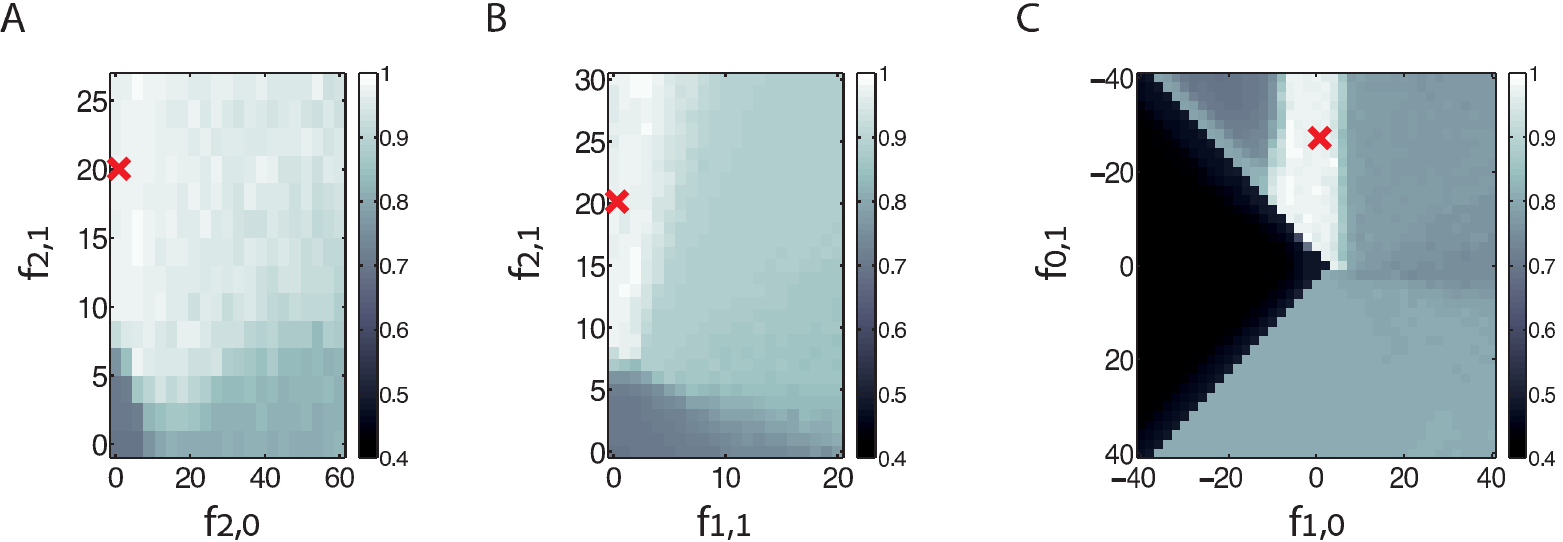}}
\end{centerline}
\protect\caption{\small \textbf{\label{FigS: second_order_motifs} Influence of second order motifs on synfire chains formation.} 
\textbf{A.}  Chain score as a function of the motif coefficients $f_{2,0}$ (horizontal axis) and $f_{2,1}$ (vertical axis).
\textbf{B.} Chain score as a function of the motif coefficients $f_{1,1}$ (horizontal axis) and $f_{2,1}$ (vertical axis).
The simulations in A and B include also a contribution from the 
motif $\{0,1\}$ with fixed $f_{0,1}$, and $f_{1,2} = -f_{2,1}$.
\textbf{C.} Reproduction of Fig~\ref{fig:Synfire-chains motifs}A. The red cross designates a set of motif coefficients identical to those
marked by red crosses in panels A-B.
In all panels, each data point represents an average over ten simulations, each with a different realization of the initial random connectivity.
}
\end{figure}

\begin{figure}[b]
\begin{centerline}
{\includegraphics[scale=0.8]{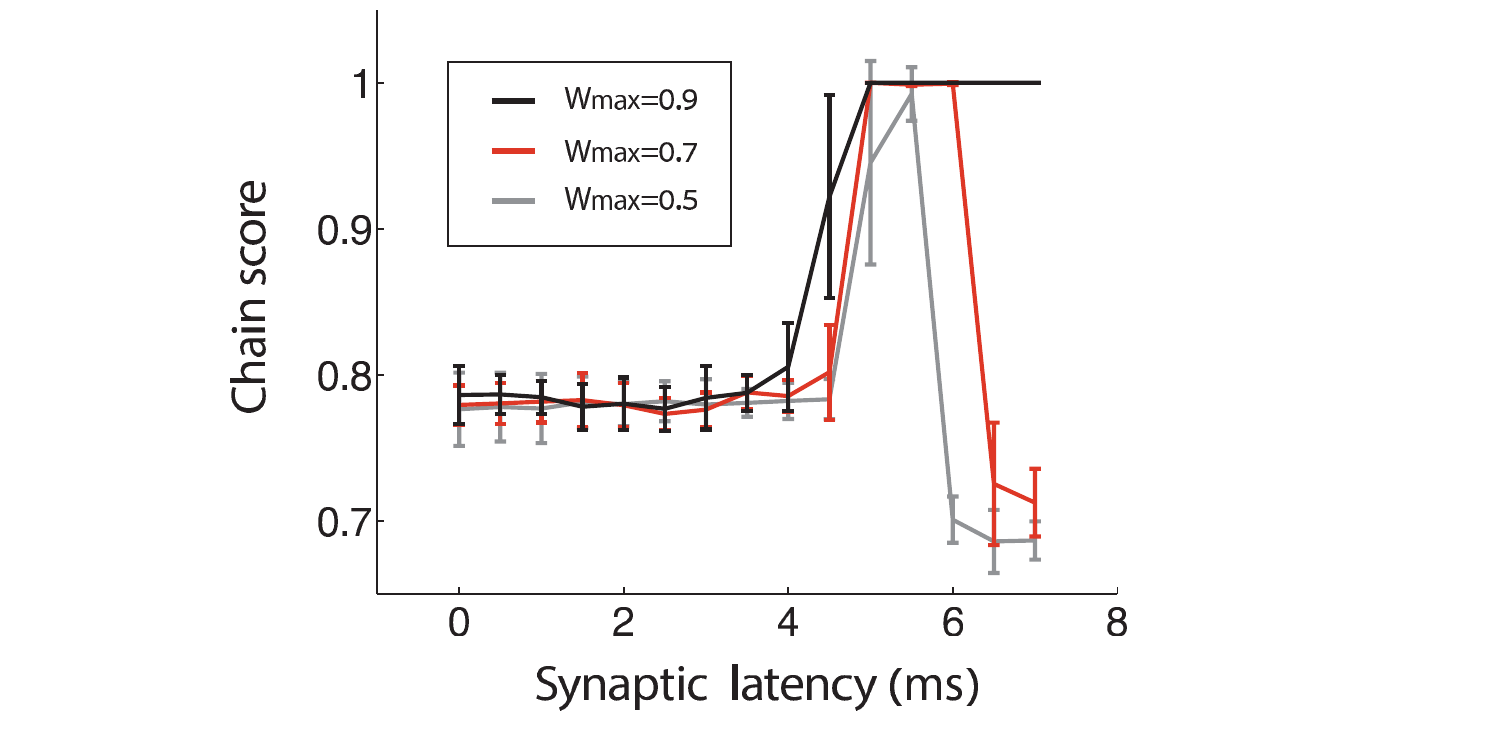}}
\end{centerline}
\protect\caption{\small \textbf{\label{S_different_W_max} Influence of $W_{\rm max}$ on synfire chain formation. } Chain score of the steady state connectivity, obtained from simulations of the complete dynamics (Eq.~\ref{eq: Presize STDP}). Horizontal axis: synaptic latency. Each line corresponds to simulations with a different
choice of $W_{\rm max}$: $0.9$ (black), $0.7$ (red), $0.5$ (gray). Each data point represents an average over ten simulations, each with a different realization of the initial random connectivity, and the
error bars represent the standard deviation of the chain score. All other parameters are specified in \textit{Methods}.
}
\end{figure}

\begin{figure}[b]
\begin{centerline}
{\includegraphics[scale=0.9]{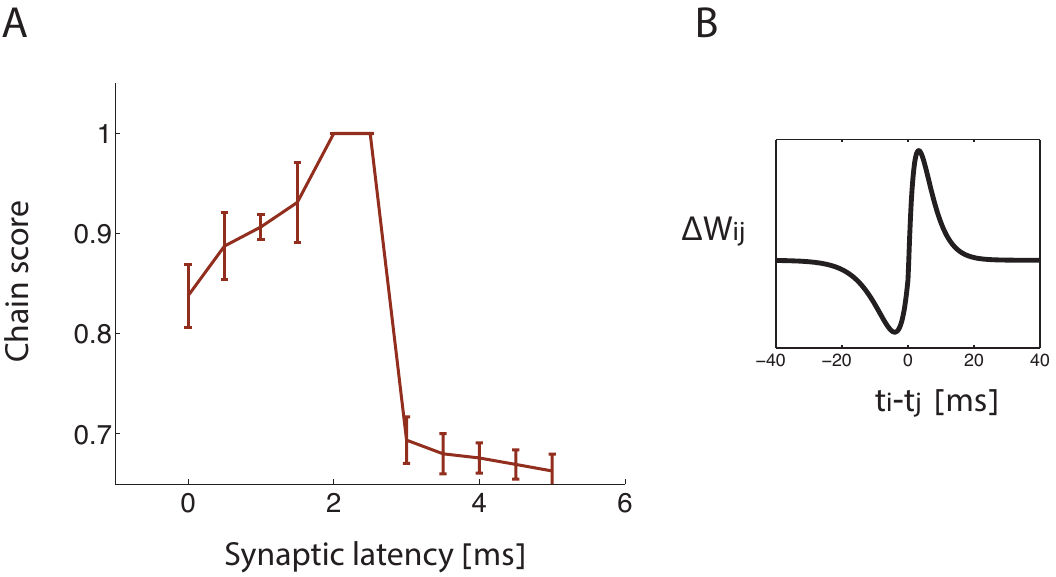}}
\end{centerline}
\protect\caption{\small \textbf{\label{S4_Fig} Formation of synfire chains under non antisymmetric STDP dynamics. A. }{
Chain score of the steady state connectivity, obtained from simulations of the complete dynamics (Eq.~\ref{eq: Presize STDP}) with a non antisymmetric STDP function (see \textit{Methods}). Horizontal axis: synaptic latency.  Each data point represents an average over ten simulations, each with a different realization of the initial random connectivity.
}\textbf{B.}{
An illustration of the non antisymmetric STDP function.
}}
\end{figure}

\begin{figure}[b]
\begin{centerline}
{\includegraphics[scale=0.75]{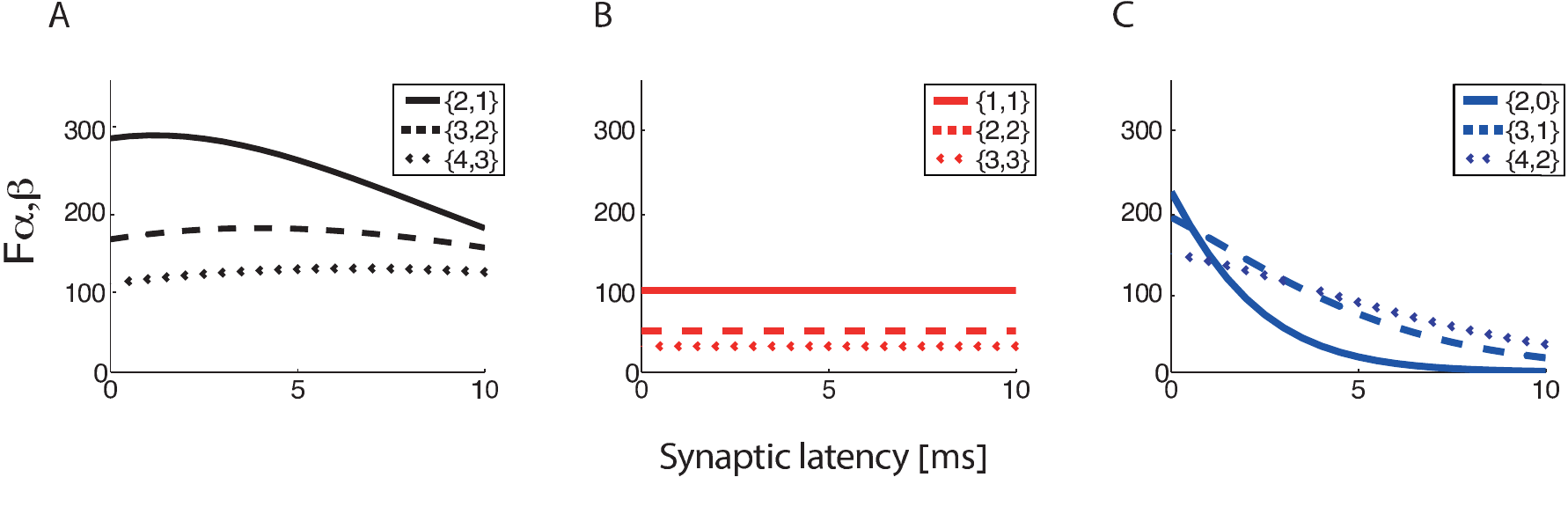}}
\end{centerline}
\protect\caption{\small \textbf{\label{S5_Fig} Contribution of high order motifs as a function of the synaptic latency. A.} {Contribution of motifs with $\alpha-\beta=1$ weakly depends on the synaptic latency.
}\textbf{B.}{
Contribution of motifs with $\alpha=\beta$ does not depend on the synaptic latency.
}\textbf{C.}{
Contribution of motifs with $\alpha-\beta=2$ decays rapidly as a function of the synaptic latency. All the parameters are specified in \textit{Methods}.
}}
\end{figure}

\nolinenumbers

%
%
%



\end{document}